\documentclass[sigconf]{acmart}

\usepackage[utf8]{inputenc}
\usepackage{algorithm}
\usepackage{algorithmic}

\usepackage{tabularx}

\newcommand{\ASERevision}[1]{\textcolor{black}{ #1}}
\usepackage{makecell}
\usepackage[export]{adjustbox}
\usepackage{xcolor}
\usepackage{float}
\usepackage{color}
\usepackage{tabularx}
\usepackage{numprint}
\usepackage{paralist}
\usepackage{balance}
\usepackage{amsmath}
\usepackage{subcaption}
\usepackage{tablefootnote}

\usepackage{hyperref}
\usepackage{graphicx,multirow}
\usepackage{newfloat,caption,subcaption,listings,mdframed}
\usepackage[font=small]{caption}
\usepackage{xspace}
\newcommand{\approach}{Reward\-Repair\xspace}

\usepackage{color}
\definecolor{bluekeywords}{rgb}{0.13,0.13,1}
\definecolor{greencomments}{rgb}{0,0.55,0.2}
\definecolor{redstrings}{rgb}{0.9,0,0}

\newcolumntype{b}{X}
\newcolumntype{s}{>{\hsize=.5\hsize}X}

\DeclareFloatingEnvironment[fileext=frm,placement={!ht},name=Listing]{listing}
\begin{document}
\title{Neural Program Repair with Execution-based Backpropagation}

\author{He Ye}
\email{heye@kth.se}
\affiliation{
  \institution{KTH Royal Institute of Technology}
  \country{Sweden}
}

\author{Matias Martinez}
\email{matias.martinez@uphf.fr}
\affiliation{%
  \institution{Université Polytechnique Hauts-de-France, France \& KTH Royal Institute of Technology, Sweden }
 \country{}
}

\author{Martin Monperrus}
\email{martin.monperrus@4open.science}
\affiliation{%
  \institution{KTH Royal Institute of Technology}
  \country{Sweden}
}



\begin{abstract}

Neural machine translation (NMT) architectures  have achieved promising results  for  automatic program repair.
Yet, they  have the limitation of generating
low-quality  patches (e.g., not compilable patches). 
This is because the existing works only optimize a purely syntactic loss function based on characters and tokens without incorporating program-specific information during neural network weight optimization. 
In this paper, we propose a novel program repair model called RewardRepair. 
The core novelty of RewardRepair is to improve NMT-based program repair with a loss function based on program compilation and test execution information, rewarding the network to produce patches that compile and that do not overfit.
We conduct several experiments to evaluate RewardRepair showing that it is feasible and effective to use compilation and test execution results to optimize the underlying neural repair model. RewardRepair correctly repairs 207 bugs over four benchmarks. 
we report on repair success for 121 bugs that are fixed for the first time in the literature.
Also, RewardRepair produces up to 45.3\% of compilable patches, an improvement over the 39\%  by the state-of-the-art.
\end{abstract}

\maketitle

\section{Introduction}


Automatic program repair (APR) aims to reduce manual work related to bug localization and bug fixing~\cite{Monperrus2015,TSE-repair-survey}. With recent advances in deep learning, research has been proposed to use neural networks for program repair, a subarea of a trend on using machine learning on code \cite{machineonlearningoncodesurvey}. This line of work, put here under the umbrella term ``neural program repair'', mostly uses neural machine translation (NMT) approaches \cite{CURE-icse21,SEQUENCER,CoCoNuT,Tufano-ICSE19,codit-tse20,Tufano-tse19,deepfix}. 

Program repair systems based on neural machine translation treat the repair task as a translation from buggy code to correct code, both represented as a sequence of tokens~\cite{SEQUENCER}. 
Given sufficient training data, NMT-based repair has achieved promising
performance \cite{CURE-icse21,SEQUENCER,CoCoNuT,Tufano-ICSE19,codit-tse20,Tufano-tse19,deepfix}. 
All the prior works on program repair based on neural machine translation use the static loss function: cross-entropy loss based on token similarity. 


Despite early successes, NMT-based program repair approaches suffer from two major drawbacks.  First,
they often generate patches that do not compile \cite{SEQUENCER}. The reason is that a lower cross-entropy value does not necessarily lead to a compilable patch.
Second, the loss function under optimization forces the neural network to learn to produce absolutely identical patches, thus missing the opportunity to explore semantically equivalent.
However, Rabin et al. \cite{rabin2021generalizability} made the case that generalizability in machine learning on code relates to generalization over equivalent programs.
Both problems (uncompilability and being stuck with syntactic identity) share the same root: there is a discrepancy between the core training objective,  learning to generate compilable and correct patches, and the loss function that is being optimized.
The cross-entropy loss used in related work requires a strict pairwise matching between the generated patch and the human-written ground truth patch, and not more~\cite{NMT-bridging}. Nothing in cross-entropy loss encourages the neural network to produce compilable or syntactically different but semantically equivalent patches. 
This is the problem we address in this paper.

We introduce a novel neural repair model, called \approach based on a mixed learning objective. The key insight in the design of \approach is the combination of a syntactic training objective at the token-level and a semantic training objective based on program execution.  
\approach defines a discriminator to discriminate good patches from low-quality ones during the training of the neural network. The discriminator is based on executing the compiler and running the test cases on the generated patches, providing high qualified execution feedback on their quality. 
This feedback is transformed as a quantified reward signal that modulates the cross-entropy loss.
Then, the neural network's weights are updated based on this novel discriminator. In other words, in \approach, backpropagation embeds essential compilation and execution information. 

We conduct large experiments to evaluate \approach based on four well-accepted datasets from the literature, including Defects4J version 1.2~\cite{defects4j}, Defects4J  version 2.0~\cite{defects4j}, Bugs.jar~\cite{Bugsjar-MSR18} and QuixBugs~\cite{lin2017quixbugs}.
First, we show that RewardRepair produces more correct patches than the recent related work. In total, RewardRepair repairs 207 on the four benchmarks.
RewardRepair achieves an improvement in two benchmarks Defects4J(v2.0) and Bugs.jar, and achieves the top-2 performance in the other two benchmarks Defects4J(v1.2) and QuixBugs. 
Second, we demonstrate that RewardRepair outperforms the state-of-the-art on addressing the compilability problem in neural program repair, CURE \cite{CURE-icse21}, by producing a higher ratio of compilable patches in all considered beam size configuration (45.3\% versus 39\% in top-30, and 37.5\% versus 28\% in top-100 candidate patches).

To sum up, our contributions are:
\begin{itemize}
    \item We devise, \approach, a neural program repair approach with execution-based backpropagation. \approach defines a novel training objective that employs compilation and test execution information to optimize the neural network.
    \item We perform an original series of experiments to show that  \approach's training objective outperforms the cross-entropy loss used in related work. Our experimental results demonstrate that embedding execution information in backpropagation  improves the quality of generated patches (more compilable patches and correct patches).
    \item We provide evidence that \approach can correctly fix 45 bugs for Defects4J(v1.2), 45 bugs for Defects4J(v2.0), 97 bugs for Bugs.jar and 20 bugs for QuixBugs. We report that ReparRepair can correctly fix 121 bugs that were never repaired in previous literature, including 5 unique Defects4J(v1.2) bugs.
    \item We make all our data and code publicly available for future research \cite{experiment}.
\end{itemize}

\section{Background}
\label{sec:background_neural}

\subsection{Neural Program Repair}

Neural Machine Translation (NMT) systems have recently achieved  state-of-the-art performance on program repair tasks, forming a field called ``neural program repair'' \cite{SEQUENCER,CoCoNuT,codit-tse20,Tufano-ICSE19,deepfix,Tufano-tse19, Recoder,CURE-icse21}.
Despite the difference in their inputs and neural models, those works are  similar in the sense that they are all based on
a typical NMT model formulated as an encoder-decoder-attention architecture optimized with cross-entropy loss function \cite{nmt,attention-all-you-need,Vaswani2021Transformer,NMT-bridging,overcorrection}. 
All the prior works on program repair are based on the NMT architecture with a cross-entropy loss function to update the neural network weights \cite{SEQUENCER,CoCoNuT,codit-tse20,Tufano-ICSE19,deepfix,Tufano-tse19, Recoder,CURE-icse21}. 
The cross-entropy loss (\textit{a.k.a.} log loss) is a measure from information theory, building upon entropy and calculating the difference between two probability distributions \cite{nmt,aligncrossentropy}. During the training of a neural repair model, the cross-entropy loss calculates the difference between the generated tokens and the human-written patch  tokens in a strict pairwise matching manner \cite{nmt,NMT-bridging,overcorrection},
and is used to update the neural network weights by backpropagation. 
In program repair patches, a low cross-entropy value  means that the generated patch is syntactically close to the ground truth patch at the token-level.

\subsection{Limitations of Current Neural Repair}
\label{sec:background:problems}

The token-based cross-entropy loss optimization is effective at guiding the model to generate patches syntactically identical or close to the human-written patches given as input during training.
However, it suffers from two major limitations.
Firstly, a major goal in patch generation is to generate well-formed patches that compile. 
Unfortunately, the cross-entropy loss does not favor patches that compile over non-compilable patches.
Even worse, if a generated patch has a significantly lower loss per the token-based cross-entropy, although being non-compilable, it would be favored by the model at training time.
Secondly, the cross-entropy loss function fails to learn from semantically equivalent patches: a syntactically different but semantically equivalent patch could potentially have a high cross-entropy loss value. 
This means that the cross-entropy loss 
discourages the network to explore equivalent solutions. 
In the field of neural machine translation (NMT), this problem is known as the overcorrection problem of cross-entropy loss~\cite{NMT-bridging,overcorrection}: cross-entropy based models tend to learn strictly identical translations and to overcorrect synonymous tokens which would be acceptable.
Overall, the problem we address in this paper is that the cross-entropy loss used in previous research cannot learn programming knowledge beyond tokens.

\subsection{Motivating Example}
\label{sec:motivating-example}

\begin{listing}[t!]
\noindent\begin{minipage}[b]{0.5\textwidth}
    \begin{lstlisting} [firstnumber=419] 
<@\colorbox{red!30}{- return FastMath.pow(2 * FastMath.PI, -dim / 2 ) *\quad}@>           
<@\colorbox{green!30!}{+ return FastMath.pow(2 * FastMath.PI, -0.5 * dim ) *}@>
    \end{lstlisting}
 
     \subcaption{The human-written patch}
   \label{motivate-human-patch}  
    \end{minipage}%
    \hfill
    \begin{minipage}[b]{0.49\textwidth}
    \begin{lstlisting}[firstnumber=419] 
<@\colorbox{green!30!}{+ return FastMath.pow(2 * FastMath.PI, -d * dim ) * }@> //loss value 0.1157 
    \end{lstlisting}
 
    \subcaption{A generated non-compilable patch receives a smaller loss score}
 \label{motivate-noncompile-patch}   
\end{minipage}%
\hfill
    \begin{minipage}[b]{0.49\textwidth}
    \begin{lstlisting}[firstnumber=419] 
<@\colorbox{green!30!}{+ return FastMath.pow(2 * FastMath.PI, -dim / 2d ) *}@> //loss value 0.4224 
    \end{lstlisting} 
    \subcaption{A generated semantically-equivalent patch receives a bigger loss score}
    \label{motivate-correct-patch} 
\end{minipage}%
\caption{Motivating example: NMT based repair models based on cross-entropy loss may favor non-compilable patches.}
\label{lst:motivatingExample}
\end{listing}

\autoref{lst:motivatingExample} is a motivating example to show the drawbacks of a neural repair model based on cross-entropy loss optimization. 
\autoref{lst:motivatingExample}a presents the buggy code and the human-written patch for bug \textit{Math-11} from Defects4J(v1.2).
\autoref{lst:motivatingExample}b  gives one  generated non-compilable patch because of the undefined variable $d$. The network generates this patch
with maximum likelihood estimation because its cross-entropy loss value is low (\textit{0.1157}).
\autoref{lst:motivatingExample}c is a semantically equivalent patch compared to the human-written patch and its computed loss is \textit{0.4224}, which is higher than the non-compilable patch.

With cross-entropy loss, the NMT-based repair model penalizes the semantically equivalent patch and favors the non-compilable patch.  
This is because there is only one token difference for the generated non-compilable patches, where the wrong token \texttt{d} is not the expected token \texttt{0.5} from the ground truth patch.
However, the token $d$ is an undefined variable in the buggy program. 
On the contrary, there exist three token differences in the semantically equivalent patch: ($dim \rightarrow \texttt{0.5}$), ($/\rightarrow *$), ($2d \rightarrow  dim$). 
Consequently, the NMT-based repair model considers the generated 
non-compilable patch is closer to the human-written patch, and thus should be favored during backpropagation.
However, in the context of program repair, the semantically equivalent patch should have a loss close to zero, because it is a valid solution to the bug.
The generated non-compilable patch  has a  lower loss, but still cannot satisfy the compiler, which is inconsistent with our goal.
This example shows the fundamental limitation of optimizing neural networks with the traditional token-based cross-entropy loss for the program repair task.


\begin{figure*}
\includegraphics[width=0.98\textwidth,height=0.36\textheight]{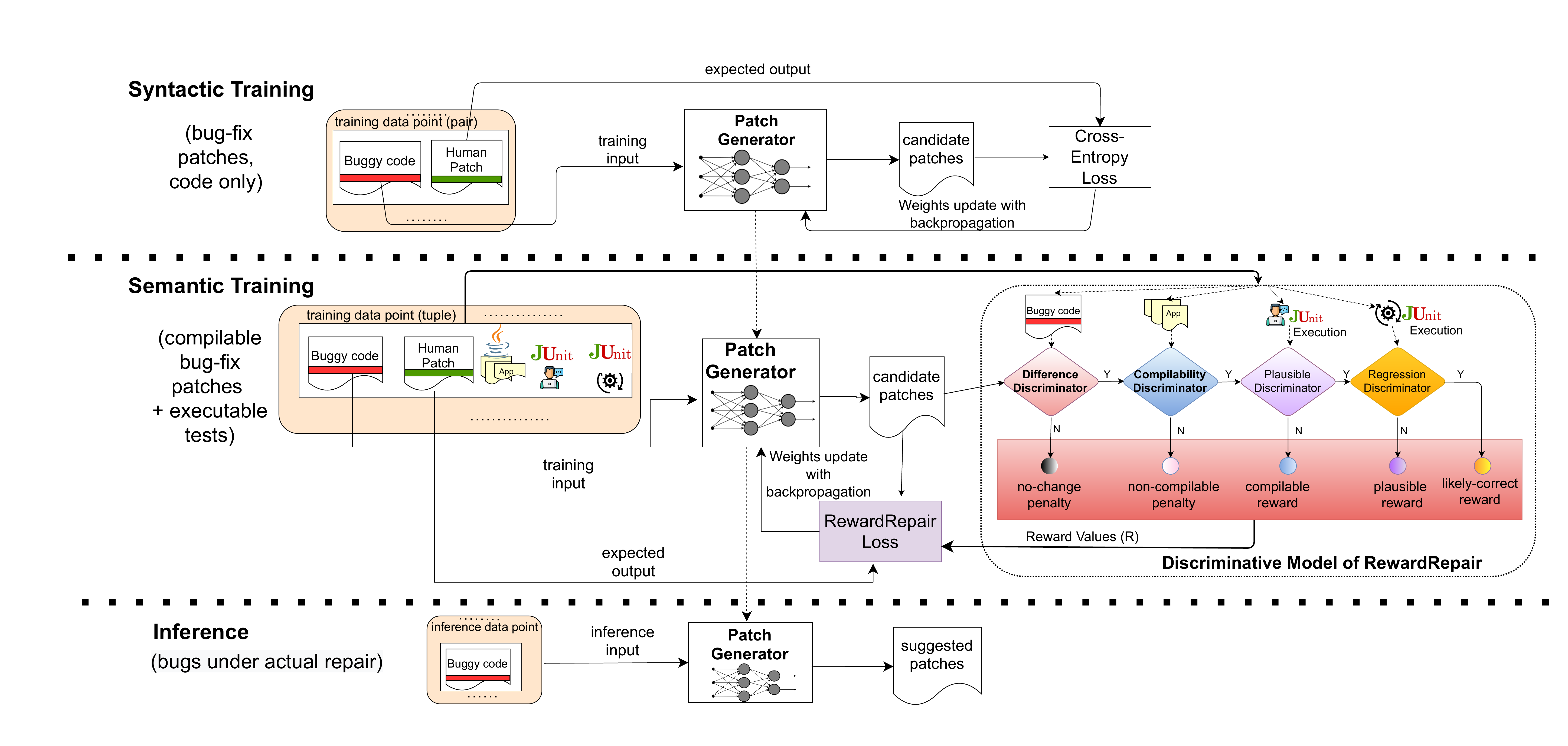} 
 \caption{An Overview of \approach. }
\label{fig:RewardRepair-overview}
\end{figure*}


\section{RewardRepair}
\label{sec:rewardrepair}


\subsection{Overview}
\label{sec:overview}

In this paper, we propose a novel neural repair model called
RewardRepair.
The core idea of RewardRepair is to improve the learning process of neural program repair and in particular to address the limitations of using the cross-entropy loss function at the token-level. 
\autoref{fig:RewardRepair-overview} gives an overview of \approach.
The top, middle and bottom parts represent three stages of RewardRepair: syntactic training, semantic training and inference. 
\approach is trained with two datasets respectively, a syntactic training dataset with pairs of buggy and patch code, and a semantic training dataset.
They are fundamentally different. The syntactic training dataset only consists of textual patches as in the related work \cite{CURE-icse21,CoCoNuT,SEQUENCER}.
However, the requirements for the semantic training dataset are full execution: each sample in the dataset comes with a compiler configuration and test cases. Achieving full execution enables us to derive execution-based information to be used for optimizing the neural network weights with programming knowledge during backpropagation.

\textit{Syntactic training} is our first training phase.
We train \approach to optimize the cross-entropy loss based on a large bug fix corpus per the related work \cite{CURE-icse21,CoCoNuT,SEQUENCER,deepfix,DLFix,Tufano-ICSE19,codit-tse20}. 
Syntactic training is meant to provide a good initial model  for semantic training. 
\textit{Semantic training} is our second phase after syntactic training.
Semantic training is based on a discriminative model.
The discriminative model of \approach analyzes the candidate patches with four discriminators (\textit{difference discriminator, compilability discriminator, plausibility discriminator and regression discriminator}), and  modulates the cross-entropy loss before the start of backpropagation on training.
The \textit{inference} is the final phase. Once \approach has been trained,  it can generate patches for new and unseen bugs based on the trained patch generator.

\subsection{Code Representation}
\label{sec:coderepresentation}
As code representation, we follow Lutellier et al. \cite{CoCoNuT} to represent the buggy code and context code as two separate token sequences.
In \approach, the context code is considered as 10 lines of code surrounding the buggy code.
In addition, the context code is enriched with a summary of the buggy class as proposed by Chen et al. \cite{SEQUENCER} as follows: we keep all the instance variables and their initializers, along with the signature of the constructors and methods. 
We follow the existing work \cite{CURE-icse21} to use subword tokenization with SentencePiece \cite{Kudo2018SentencePieceAS}, as demonstrated useful by Karampatsis et al. \cite{karampatsis2020big}.

\subsection{Patch Generator}
\label{sec:patch_generator}

In \approach, the patch generator is trained in a supervised manner based on an encoder-decoder architecture \cite{T5,attention-all-you-need}.
Syntactic training takes as input buggy code tokens $B = (b_0, b_1\ldots b_n )$, context code tokens $C = (c_0, c_1\ldots c_m )$, and tokens from the ground truth fix patch $F = (f_0, f_1\ldots f_t )$.
\approach transforms the $B$ and $C$ into a predicted patch  $F'= (f'_0, f'_1\ldots f'_k )$. 
Note that size of the buggy code, context code, ground truth patch and predicted patch, i.e., $n$, $m$, $t$ and $k$, can be different. 
In \approach, patch generation is shared by both syntactic training and semantic training.



\subsection{Syntactic Training of \approach}
\label{sec3:syntactic_training}

\approach initially trains the patch generator with cross-entropy loss function per the state-of-the-art of NMT for program repair \cite{CoCoNuT,SEQUENCER,deepfix,DLFix,Tufano-ICSE19,codit-tse20,CURE-icse21}. 
For each training point, the optimization target is to minimize the loss between ground truth fix patch $F$ and the predicted patch $F'$. 
As shown in the previous study \cite{CURE-icse21,CoCoNuT},  syntactic training is trained with a large corpus of buggy code and fixed code pairs.
Syntactic training could be trained for multiple epochs to achieve convergence and obtain the best combination of weights.
By the end of syntactic training, the patch generator's weights between the connections of the networks are optimized.

\subsection{Semantic Training of \approach}
\label{sec-discriminator}
\label{sec:semantic_training}

The goal of semantic training is to let the patch generator be aware of program-specific knowledge (compilation and execution) beyond the syntactic loss computation at the token-level. For that, we propose a mixed learning objective \cite{overcorrection}, where ``mixed'' means that the core learning objective is combined with two or more sub-learning objectives. 
Our  mixed learning objective  combines the core cross-entropy objective with compilation and execution information.
Once the \approach has been sufficiently trained with syntactic training, we start the semantic training process alternately with syntactic training.

For semantic  training,  \approach employs a discriminative model to assess the quality of generated patches based on compilation and test oracles. As a result, the discriminative model outputs a reward value that quantifies the patch quality, which is used to adjust the weights of the patch generator during backpropagation.  
A higher reward means  better quality for the generated patch, e.g., the patch is compilable and passes all test cases.   On the contrary, a lower reward means that the quality of the generated patch may be unsatisfying, e.g., the patch is non-compilable.

Precisely, the patch reward value modulates the token-level loss value before the start of the backward pass \cite{micikevicius2018mixed,lossscale,AnchorLoss}, so that the updated loss can then be properly represented with program-specific knowledge.  
In such a way, the discriminator guides the neural network to generate high-quality patches.
To our knowledge, we are the first to introduce  semantic training based on a patch discriminative model for neural program repair.

\subsubsection{Discriminative model}

In machine learning, a discriminator is a model for identifying ``good'' and ``bad'' predictions~\cite{GanGoodFellow}.
The discriminator of \approach is the key component of the semantic training phase.
Its goal is to measure the quality of the generated patches, which is then quantified as a reward signal. 
In \approach, the discriminative model is composed of four discriminators, each of them specialized in one aspect of patch quality.
These four discriminators are serially executed. 
Each discriminator does a binary classification: whether a patch fulfills the discriminator's criterion or not. When a discriminator is affirmative, it means the patch passes the quality criterion, and it is passed to the next discriminator. 
In the rest of this section, we present the four discriminators of \approach.

\paragraph{Difference discriminator} 
\label{sec:difference_discriminator}
The difference discriminator validates whether the generated patches are different from the given buggy code.  
It is required because we found that neural repair models regularly generate a patch that is identical to the given buggy code, i.e., the output of the model is identical to the input of the model (called a ``no-change'' patch in this paper). 
This happens when the neural nets discover that the buggy code achieves the maximum likelihood estimation. 
This is explained by previous research \cite{TianASE20}, which has shown that the many buggy codes are similar to correct code, with only minor transformations and few changed tokens.
Consequently, the generator tends to copy the buggy code because it is the maximum likelihood prediction, per the data seen at training time.
We design the difference discriminator to syntactically compare the generated patch code with the input buggy code, with a token sequence comparison.
If the buggy code and generated patched code are the same, the  difference discriminator assigns the generated patch a penalty, called in this paper the \textit{no-change penalty} ($R_{no-change}$), i.e., a negative reward.
This penalty signal modulates the \approach loss to avoid the generation of no-change patches.

\paragraph{Compilability discriminator} 
\label{sec:compiler_discriminator}

The compilability discriminator validates whether the generated patches are compilable. 
As shown in previous research \cite{CURE-icse21}, neural repair models suffer from outputting  non-compilable patches. 
For example, the golden sequenceR model produces 79.7\% of patches that are not compilable in top-50 candidate patches. 
To force the generator towards producing compilable patches, we introduce a compilability discriminator in the semantic training phase.
The compilability discriminator employs the compiler to compile the generated patched program.
If the patched program is compilable, \approach assigns a \textit{compilable reward} ($R_{compilable}$) and passes the patch to the next discriminator for further assessment. Otherwise, a \textit{non-compilable penalty} ($R_{non-compilable}$), i.e., a negative reward, is returned to the \approach model.

\paragraph{Plausibility  discriminator} 
\label{sec:plausible_discriminator}
The plausibility discriminator aims at encouraging the generator to produce patches that pass the human-written test cases provided by developers.
Recall that, per the seminal work on program repair \cite{LeGoues2012GenProg}, test cases can be used as an executable program specification.
In other words, if a patch passes the human-written tests, it is considered as a plausible patch. 
During semantic training, \approach is trained on buggy projects with executable human-written tests.
Each candidate patch is executed against the human-written tests.
If a patch makes all human-written tests pass, a \textit{plausible reward} ($R_{plausible}$) is assigned. 
By doing so, the plausibility discriminator leverages the human-written tests to drive the network to produce plausible  patches that pass all human-written tests.

\paragraph{Regression discriminator}
\label{sec:behavior_discriminator}
The last discriminator used in  \approach's discriminative model is the regression discriminator.
The goal of this discriminator is to minimize the behavioral changes introduced by the patch.
This discriminator complements the plausible patch discriminator by specifying behavior outside the human-written tests, in order to avoid patch overfitting \cite{zhongxing-EMSE18,Le:overfitting,le:reliability-patch-assess,CURE-worse-15}.
The \approach regression discriminator employs automatically generated tests, per the RGT technique of Ye et al. \cite{drr}.  The effectiveness of this technique to identify correct patches from plausible patches has been shown in recent work \cite{ASE20Wang,ODS,quixbugs-jss}.
The idea of the RGT technique is to automatically generate test cases based on the ground truth patched program to expose program execution behavior differences between a generated patch and a ground truth patch \cite{le:reliability-patch-assess,ASE20Wang,drr}. 
If a candidate patch makes all RGT tests pass, i.e., it does not contradict the ground truth program behavior, it is considered as likely-correct.

During semantic training, all patches are executed against the RGT tests.
If a candidate patch makes all automatically generated tests pass, meaning the same program execution behavior with ground truth program,  then  a \textit{likely-correct reward} signal ($R_{l-correct}$) is assigned to this patch. 
This discriminator's reward is used to encourage the \approach to avoid regressions.
\emph{This means we encourage \approach to generate non-overfitting patches beyond the existing test cases.}

\subsubsection{Defining reward values from discriminators}
\label{sec:reward-value}

As shown previously, \approach defines five reward signals for patch quality assessment.
The discriminators are executed serially, that is, if a patch \emph{does not satisfy} one discriminator, then the reward $R$ obtained up to that moment is returned immediately and other discriminators are not executed. 
Consequently, the $R$ is the maximum  of the five reward values as follows:

\begin{equation}
\footnotesize
R = 
max
\begin{cases}
R_{no-change} = s_0\\
R_{non-compilable} = s_1 \\
R_{compilable} = s_2   \\
R_{plausible} = s_3   \\
R_{l-correct} = s_4  \\
\end{cases}
\label{equ-reward-values}
\end{equation}

where  $s_{i}$  ($i\in \{0, 1, 2, 3, 4\}$) are five scaling parameters that control the range of the reward values.
The scaling parameters of $s_{i}$ define a configuration space that is controlled by end-users of \approach. 
Additionally, those reward values must fulfill the following constraint: 
\begin{equation}
\footnotesize
R_{no-change} <  R_{non-compilable} <  R_{compilable} < R_{plausible} <  R_{l-correct}
\end{equation}
where the higher reward value represents the better quality of the generated patch.

Given the cross-entropy $\mathcal{L}$,
the loss function of \approach  $\mathcal{L}_{RewardRepair}$ dynamically reweighs $\mathcal{L}$ with respect to the discriminator's reward signal $R$, which encodes the quality of the generated patch.
We follow \cite{seqGan-aaai17,AnchorLoss} to formulate \approach loss function as a scaling of the cross-entropy, as follows:

\begin{equation}
\footnotesize
\mathcal{L}_{RewardRepair}  = (1-R)*\mathcal{L} \quad  (where \quad R<1)
\label{loss-equ}
\end{equation}

The reward modulator $(1-R)$  constrains the domain of $R \in (-\infty, 1)$, as it is meaningless for the  objective function to minimize a negative loss. 
In this formulation, the
syntactic cross-entropy at the token-level is combined with the semantic reward at patch-level, embedding compilation and execution knowledge deep into the neural model.
It mitigates the limitations of only considering cross-entropy loss in program repair tasks and solves the problem discussed in  Section \ref{sec:motivating-example}. 

For low-quality  patches, \approach assigns a penalty, i.e., negative reward value, to increase the original cross-entropy loss  $\mathcal{L}$. Thus the domain for $R_{no-change}, R_{non-compilable} \in (-\infty, 0)$.
On the contrary, for the high-quality patches, \approach scales down the original $\mathcal{L}$ by assigning positive reward values for $R_{compilable}$, $ R_{plausible}$, and $R_{l-correct}$ to encourage the model. The domain for these three reward values is 
$[0, 1)$.
Based on \autoref{loss-equ}, the higher the reward, the smaller computed loss is back-propagated into the neural model. 
The extreme case is the highest reward value approximate to \textit{1}, where the \approach loss goes close to \textit{0}. This indicates that the generated patch is likely correct and the network should not be changed.

Combining the domain constraints discussed above, 
the range of scaling parameters $s_{i}$ must meet the following criteria:

\begin{equation}
\footnotesize
\begin{cases}
s_0 \in (-\infty,0) \\
s_1 \in (s_0, 0) \\
s_i \in [0,1),\xspace s_{i-1} < s_{i}, i \in \{2,3,4\} \\
\end{cases}
\label{scale-values}
\end{equation}

\begin{algorithm}[t]
\footnotesize
  \caption{One step of semantic training in RewardRepair}
  \begin{algorithmic}[1]
  \STATE \textbf{Input:}  buggy code b, context code c, ground-truth patch code p, human-written test cases $t_h$,
   RGT test cases $t_m$,
  learning rate $\alpha$, G is the \approach patch generator, D is the discriminator function
  \STATE $\widetilde{b} \gets \varphi (b)$ \{Encode buggy code\}\label{algo1:buggycode}
  \STATE $\widetilde{c} \gets \varphi (c)$ \{Encode context code\} \label{algo1:contextcode}
  \STATE $\widetilde{p} \gets \varphi (p)$ \{Encode ground-truth patch code\} \label{algo1:gtcode}
  \STATE $ \widetilde{q} \gets G (\widetilde{b}, \widetilde{c})$ \{Generate candidate patch \}\label{algo1:generate} 
  \STATE $\mathcal{L} =  \sum_{x \in X } \widetilde{p}(x)  log \,  \widetilde{q}(x) $ \label{algo1:celoss}
  \STATE $ R \gets D (\widetilde{b},\widetilde{c},\widetilde{q}, t_h, t_m)$ \label{algo1:discrim}
  \STATE $\mathcal{L}_{\approach}  = (1-R)* \mathcal{L} $ \label{algo1:semanticloss}
  \STATE $G \gets G -  \alpha \partial L_{\approach}/\partial G  $ \{update patch generator with backpropagation\} \label{algo1:updategenerator}
  \end{algorithmic}
  \label{alg:SemanticRewardRepair}
\end{algorithm}


\subsubsection{Algorithm}


Algorithm~\autoref{alg:SemanticRewardRepair}  presents one step of semantic training. 
Given the encoded buggy $\widetilde{b}$ and context code $\widetilde{c}$ (line \ref{algo1:buggycode} and \ref{algo1:contextcode}), the patch generator $G$ generates a candidate patch (line \ref{algo1:generate}).
Then, the cross-entropy loss $\mathcal{L}$ is computed by comparing the token distribution between the ground truth patch $\widetilde{p}$ and the generated patch $ \widetilde{q}$ (line \ref{algo1:celoss}), where the $x$ indicates the index of tokens.
The discriminator provides a reward $R$ based on generated patch  $ \widetilde{q}$ and the corresponding program compilability and test execution information (line \ref{algo1:discrim}). 
Lastly, \approach combines the cross-entropy loss at token-level  and reward value at patch-level  to form \approach loss (line \ref{algo1:semanticloss}), which encodes the program-specific knowledge.

\subsection{Inference}
\label{sec:inferencephase}

At inference phase, for a given suspicious statement found by fault localization tools (e.g., Ochiai \cite{fl-tool}), \approach{} represents it with two sequences of tokens: one for the suspicious statement, the other one for its context (see Section \ref{sec:patch_generator}).
Those tokens are given to the patch generator of \approach{}, previously trained as explained in Sections~ \ref{sec3:syntactic_training}~and~\ref{sec:semantic_training}.
As  \approach is configured by the inference beam size $n$ (see \cite{CURE-icse21,SEQUENCER}), it outputs  the $n$ best patches for that suspicious statement.
RewardRepair can be used with any fault localization technique in  real-world bug repair tasks, as shown by \cite{Rhero}.

\subsection{Implementation}
\label{sec-implementation}

We implement \approach's patch generator  with the state-of-the-art Transformer based architecture \cite{T5} from Hugging Face. 
RewardRepair is trained with 15 syntactic training epochs and 4 semantic training epochs. 
For hyper-parameters configuration, we use a vocabulary size of \numprint{32128}.
We configure \approach to take a maximum of $512$ input tokens from buggy and context code, and generate a patch with a maximum of $100$ tokens. 
The learning rate sets to $1\mathrm{e}{-4}$ for both syntactic and semantic training. We configure reward scaling values $s_{i}, i \in \{0,1,2,3,4\}$ respectively to \{-0.4, -0.2, 0.2, 0.4, 0.6\} for the best experiment result. 
The encoder and decoder consist of 6 layers.
 \approach is configured by a beam size of $200$ and outputs the $200$ best patches per bug. We consider the beam size of $200$ rather than $1000$ used in CURE \cite{CURE-icse21} and CoCoNuT \cite{CoCoNuT} due to the limitations of our available GPUs.


\section{Experimental Methodology}

In this section, we describe our methodology for evaluating \approach by defining three research questions and how we propose to answer them.

\subsection{Research Questions}

\begin{itemize}

\item RQ1 (comparison with other tools): To what extent is RewardRepair effective at repairing bugs compared with the state-of-the-art repair approaches?

\item RQ2 (compilable rate): To what extent does RewardRepair improve the compilability of generated patches? 

\item RQ3 (impact of semantic training): To what extent does semantic training improve the effectiveness of RewardRepair?

\end{itemize}

\subsection{Dataset}
\label{sec:datasets}
\begin{table}[t]
\footnotesize
\renewcommand{\arraystretch}{1.28}
\begin{tabular}{p{0.1\linewidth}p{0.35\linewidth}p{0.28\linewidth}p{0.1\linewidth}}

\hline
 \bf{Phases} & \bf{Requirements} & \bf{Name \& Source}  & \bf{\#Patches} \\
\hline
 \multirow{3}{*}{\makecell{Syntactic\\training}} & \multirow{3}{*}{Tokenization}   &CoCoNuT \cite{CoCoNuT} & \numprint{3241966}\\

 &&MegaDiff \cite{monperrus2021megadiff}  & \numprint{240306} \\
  &&CodRep \cite{Chen2018Coderep}  & \numprint{24969} \\

\hline
\multirow{2}{*}{\makecell{Semantic\\training}}  &\footnotesize Tokenization, Compilation &\multirow{2}{*}{Bears \cite{Bears}} & \multirow{2}{*}{123} \\
& \footnotesize Developer and RGT Tests&&\\
\hline

\multirow{4}{*}{Testing} & \multirow{4}{*}{ \makecell{\footnotesize Tokenization, \footnotesize Compilation,\\ Developer Tests}} & Defects4J(v1.2) \cite{defects4j} & 120 \\
&& Defects4J(v2.0) \cite{defects4j} & 257 \\
&& Bugs.jar \cite{Bugsjar-MSR18} & 490 \\
&&QuixBugs \cite{lin2017quixbugs} & 34\\

\hline

\end{tabular}
\caption{Datasets used for the different steps of our experiment.}
\label{tab:dataset}
\end{table}

Recall that we need three datasets for syntactic training,  semantic training and testing. 

We have common criteria for both training and testing datasets: 
\begin{inparaenum}[1)]
\item all datasets are composed of bug-fix patches;
\item we focus on single-file patches, per previous work \cite{SEQUENCER,DLFix,tbar};
\item we focus on single-hunk patches, where the patch is confined to a single contiguous chunk of code, at a single location, per previous work \cite{SEQUENCER,DLFix,tbar}; 
\item  we discard patches that do not make program behavior differences, e.g., those with only changes in comments or logging.
\end{inparaenum}

Next, we have specific requirements per dataset.
\autoref{tab:dataset} shows the dataset of patches that we use for training and evaluating \approach.  
The first column indicates the phase where each dataset is used. 
The second column gives the requirements for each dataset. 
The third column gives the source of the dataset and the fourth column indicates the number of patches in this dataset.
For example, as shown in the first row, \approach is syntactically trained with data from three different sources, CoCoNuT \cite{CoCoNuT}, Megadiff \cite{monperrus2021megadiff} and CodRep \cite{Chen2018Coderep}.

As aforementioned in Section \ref{sec:rewardrepair}, 
the selection criteria for semantic training dataset are:
\begin{inparaenum}[1)]
\item to be able to compile the patched program;
\item to run the test cases on the patched program;
\item to be able to automatically generate tests to specify the expected program behavior. 
\end{inparaenum}
All criteria are met for 123 single-hunk bugs of the Bears dataset~\cite{Bears}, for which some available RGT tests were generated by previous research~\cite{ODS}. 
We use the evaluation mode of \approach to create more semantic training points based on beam search, as done in \cite{kommrusch2021selfsupervised}.
It is to be noted that those criteria are very strong, and neither CoCoNuT \cite{CoCoNuT}, MegaDiff \cite{monperrus2021megadiff} nor CodRep \cite{Chen2018Coderep} meets them, in particular, the patched program cannot be compiled.

To test \approach, we  use well-accepted datasets from program repair research \cite{Liu2020Efficiency, Martinez2017experiment, Durieux:2019:RepairThemAll, quixbugs-jss}:
Defects4J~\cite{defects4j}, 
Bugs.jar \cite{Bugsjar-MSR18}, and QuixBugs \cite{lin2017quixbugs}.
For all those bug datasets, the requirements of compilation and test execution are met.
In line with the most recent work \cite{Recoder}, we also consider the bugs of Defects4J version 2.0.
After filtering single-hunk bugs, we use 120 bugs from Defects4J(v1.2) and 257 additional new bugs from Defects4J version 2.0  denoted as Defects4J(v2.0) in our paper.
We use the same single-hunk bugs criteria for Bugs.jar and QuixBugs.

\subsection{Methodology for RQ1}
\label{sec:method_rq1}
In RQ1, we compare 
RewardRepair against the state-of-the-art neural repair approaches: CURE \cite{CURE-icse21}, Recoder \cite{Recoder}, CoCoNuT~\cite{CoCoNuT}, and other approaches \cite{hercules,tbar,Simfix:2018,capgen-ICSE18,Yuan2017ARJAAR,sharpFix,elixir,nopol,astor}.
Per  previous studies \cite{CoCoNuT,CURE-icse21}, we take the quantitative results from the literature.
\ASERevision{
We run RewardRepair under two fault localization modes.
First, we use spectrum-based fault localization with Gzoltar \cite{GZoltar} per previous work \cite{LeGoues2012GenProg,nopol}.
Second, we assume that the fault has been localized, an evaluation technique known as perfect fault localization and extensively used in recent work \cite{Liu2020Efficiency,CURE-icse21,CoCoNuT}.}

We compute the two traditional APR performance metrics for each testing dataset:
\begin{inparaenum}[\it 1)]
\item  the number of bugs that are correctly repaired. In our paper, a patch is deemed correctly repaired if it meets either of the two following criteria:  it is identical to the developer patch, or it is considered as correct by manual analysis done by at least two authors;
\item the number of bugs that can be uniquely repaired by individual repair approaches.
\end{inparaenum}

\subsection{Methodology for RQ2}
In RQ2, we calculate the compilable rate of RewardRepair.
As compilable rates were reported in SequenceR~\cite{SEQUENCER}, CoCoNuT~\cite{CoCoNuT} and CURE~\cite{CURE-icse21}, we use the same benchmarks Defects4J(v1.2) and QuixBugs as they do, and compare against the numbers reported in the original papers. 
We also follow the existing work \cite{CURE-icse21} and compute the compilable rate depending on the beam size.
We report on beam sizes in 30, 100 and 200.
We do not consider larger beams due to the limitations of our available GPUs. 

\subsection{Methodology for RQ3}
In RQ3, we conduct an ablation study with the goal of measuring the effect of semantic training.
To understand the contribution of semantic training, we compare the effectiveness of  \approach considering:
\begin{inparaenum}[\it 1)]
\item only syntactic training; 
\item both syntactic and semantic training.
\end{inparaenum}
For doing this study, we apply the same protocol as the one used for responding to the RQ1.
We conduct manual analysis on the unique bugs that are only repaired by including semantic training and present the most interesting categories on repair action changes.

\section{Experimental Results}

\begin{table}[t!]
\footnotesize
\renewcommand{\arraystretch}{1.28}
\begin{tabular}{lccrr}
\hline
 \textbf{Approaches} &\textbf{D4J(v1.2)} & \textbf{D4J(v2.0)} &
\textbf{Bugs.jar}& \textbf{QuixBugs} \\

& 120 bugs& 257 bugs & 490 bugs & 34 bugs   \\
\hline

 \multicolumn{5}{c}{Using Spectrum-based Fault Localization}  \\
\hline
jGenProg \cite{astor} & 5 & -& - & 1 \\
Nopol \cite{nopol} & 5& - & - & 2  \\
Elixir \cite{elixir} &26 & -&22 & -  \\
sharpFix \cite{sharpFix} &27& -&15&- \\
SimFix \cite{Simfix:2018}& 27 & 2 &- & -\\
Hercules \cite{hercules} &33& - & - &-\\
Recoder \cite{Recoder}&39 & 19 &-&17 \\
RewardRepair (this paper) & 29 & 24 & 42 & 19\\
\hline

 \multicolumn{5}{c}{Assuming Perfectly Localized Fault} \\
 \hline
SequenceR \cite{SEQUENCER} &14 & -& - & -  \\
DLFix \cite{DLFix} & 33 &- & -&- \\
TBar \cite{Liu2020Efficiency}&33 & 8 & - & -   \\

CoCoNuT \cite{CoCoNuT} &33& -&-&12 \\
CURE \cite{CURE-icse21}&45&-&-& \textbf{24} \\
Recoder \cite{Recoder}&\textbf{52}& - &-&- \\
\hline
RewardRepair (this paper) &45  & \textbf{45} &\textbf{97}&20\\
RewardRepair  Unique & 5 & 34 &78&4\\
\hline
\end{tabular}
\caption{Comparison of RewardRepair against the related work, the numbers from the related work are filtered by single-hunk bugs. Across all benchmarks, RewardRepair correctly fixes 207 bugs  and uniquely fixes 121 ones. We use 4 testing benchmarks to maximize generalizability.}
\label{tab:comparison-sota}
\end{table}

\subsection{RQ1: Comparative Study with Other Repair Approaches}
\label{sec:result_comparison}

\begin{listing}[t!]
\noindent    \begin{lstlisting} [firstnumber=419] 
if (_dataFormatReaders != null) {
  return _detectBindAndReadValues(_dataFormatReaders.findFormat(
     src,offset,length), false);}
<@\colorbox{red!30}{-  return \_bindAndReadValues(\_considerFilter(\_parserFactory.createParser(src),  }@>        
<@\colorbox{green!30!}{+  return \_bindAndReadValues(\_considerFilter(\_parserFactory.createParser \quad\quad\quad }@>
<@\colorbox{green!30!}{+\quad\quad (src,offset,length), \quad\quad\quad\quad \quad\quad \quad\quad\quad\quad\quad\quad\quad \quad\quad\quad \quad\quad \quad\quad \quad\quad\quad\quad \quad\quad }@>
   true));
    \end{lstlisting}

\caption{RewardRepair correct patch for Defects4J(v2.0) JacksonDatabind\_57}
\label{lst:JacksonDataBind-unique-patches}
\end{listing}


\autoref{tab:comparison-sota} shows the patch generation results of RewardRepair and  12 other APR approaches on four benchmarks: the two versions of Defects4J, Bugs.jar and QuixBugs.
The numbers are the correctly repaired bugs by each APR approach.
The results are those reported in the literature, by the authors of the tool or by subsequent comparative experiments \cite{Liu2020Efficiency,quixbugs-jss}. A ‘-’ indicates that the APR approach has not been evaluated on the considered benchmark, to the best of our knowledge.
\ASERevision{
Note that the first seven APR approaches were executed with spectrum-based fault localization (FL), where the later approaches assumed perfect FL. We measure RewardRepair's effectiveness with both spectrum-based and perfect FL. Next, we focus on the comparison under perfect FL as the most state-of-the-art techniques only report effectiveness with perfect FL.}


\textbf{Repaired bugs.} Overall,  RewardRepair is able to correctly repair 45 of 120 bugs on Defects4J(v1.2), 45 of 257 bugs on Defects4J(v2.0), 97 of 490 bugs on Bugs.jar and 20 of 34 bugs on QuixBugs benchmark. From these results, we make the following observations.

RewardRepair outperforms all APR approaches in two benchmarks: Defects4J(v2.0) and Bugs.jar.  
RewardRepair sets new baselines of repaired bugs for these two benchmarks. 
While the majority of APR papers showcase bugs from version 1.0 or 1.2 of Defects4J, \autoref{lst:JacksonDataBind-unique-patches} gives the correct RewardRepair patch for a Defects4J(v2.0) bug: \textit{JacksonDatabind\_57}. 
As shown, RewardRepair succeeds in reusing the surrounding variables \texttt{offset} and \texttt{length} to construct the parameter list for overridden method $createParser$. 
Recoder \cite{Recoder}, which is so far the best tool evaluated on Defects4J(v2.0) fails at repairing this bug.

\begin{figure}
         \includegraphics[width=0.338\textwidth]{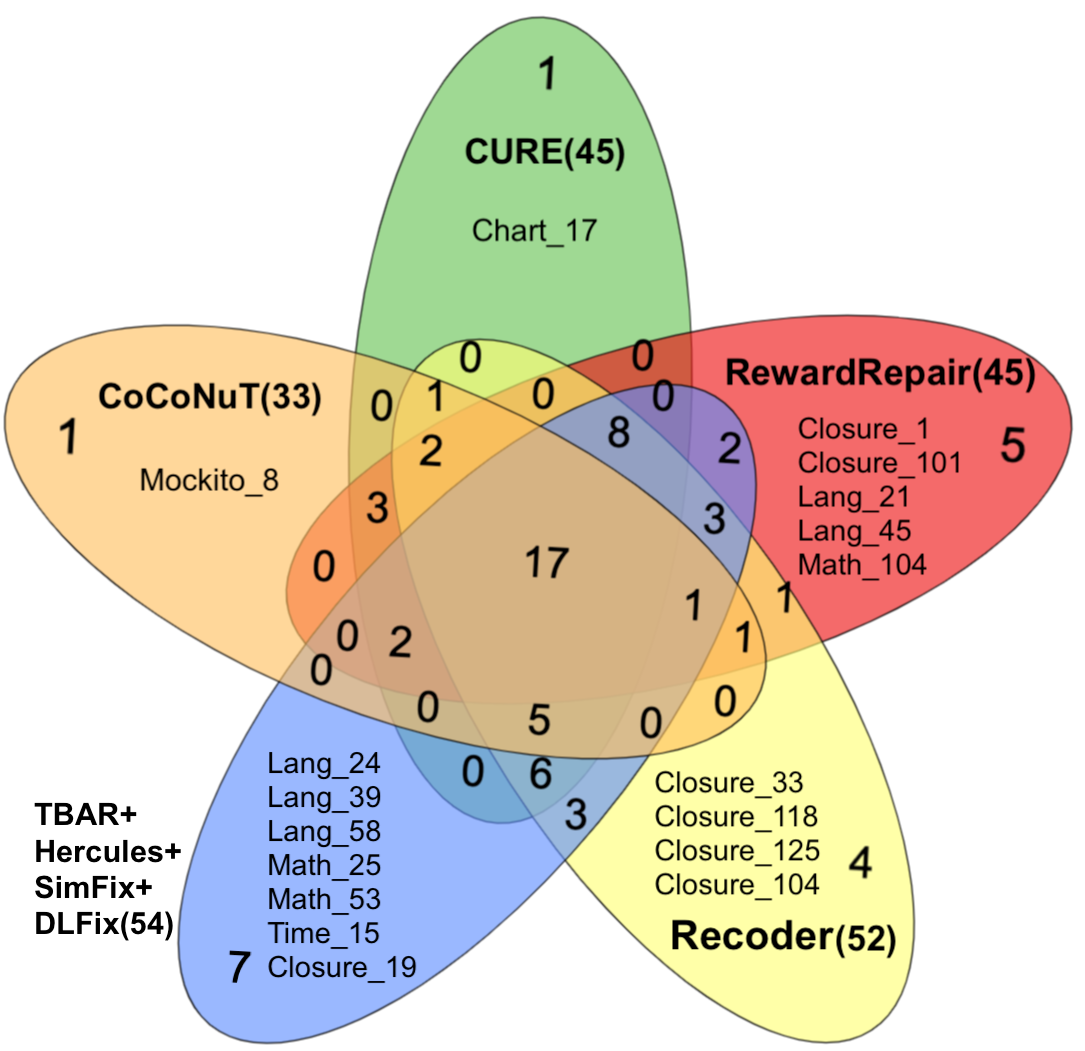}
         \caption{Uniquely repaired bugs on Defects4J(v1.2).}
         \label{fig:venn-compare}
\end{figure}

\begin{listing}[t!]
\noindent\begin{minipage}[b]{0.49\textwidth}
    \begin{lstlisting} [firstnumber=419] 
private void removeUnreferencedFunctionArgs(...){          
<@\colorbox{green!30!}{+    if (!removeGlobals) \{ \quad\quad\quad\quad \quad\quad\quad\quad\quad \quad\quad\quad\quad\quad \quad\quad\quad\quad\quad \quad\quad\quad\quad\quad \quad  }@>
<@\colorbox{green!30!}{+  \quad   return; \quad\quad\quad\quad \quad\quad\quad\quad\quad \quad\quad\quad\quad\quad \quad\quad\quad\quad\quad \quad\quad\quad\quad\quad \quad\quad\quad \quad\quad\quad\quad}@>
<@\colorbox{green!30!}{+    \} \quad\quad\quad\quad \quad\quad\quad\quad\quad \quad\quad\quad\quad\quad \quad\quad\quad\quad\quad \quad\quad\quad\quad\quad \quad\quad\quad\quad\quad \quad\quad\quad\quad \quad\quad }@>
    \end{lstlisting}

\end{minipage}%
\caption{\approach patch for Closure-1 from Defects4J(v1.2), identical to the developer patch.}
\label{lst:Closure1}
\end{listing}

In addition, RewardRepair achieves the top-2 performance on Defects4J(v1.2) and QuixBugs benchmarks.
RewardRepair performs better than all APR approaches on Defects4J(v1.2) but Recoder. 
Regarding Recoder's performance on Defects4J(v1.2), we analyze the bugs that cannot be repaired by RewardRepair. 
We find that three bugs (\textit{Closure-14}, \textit{Closure-104}, and \textit{Closure-118}) require fixing tokens outside the considered buggy class  and three bugs (\textit{Lang-26}, \textit{Lang-43} and \textit{Closure-33}) require fixing tokens outside the context code scope as implemented by RewardRepair. 
This analysis suggests that RewardRepair could achieve  better performance by enlarging the context code scope.
Regarding CURE's performance on QuixBugs, the reason is likely the beam size: recall that CURE generates 10,000 candidate patches for each bug, while RewardRepair generates 200 patches per bug. 
As shown in previous research \cite{Tufano-ICSE19}, a larger beam size leads to more correct patches.


\textbf{Uniquely repaired bugs.}
Let us now focus on the last row of \autoref{tab:comparison-sota}, which gives the number of uniquely repaired bugs by RewardRepair. 
RewardRepair respectively repairs 5, 34, 78 and 4 unique bugs for Defects4J(v1.2), Defects4J(v2.0), Bugs.jar and QuixBugs, all of which were never repaired by any other APR approaches in the literature. 
This shows RewardRepair complements all the existing works on the four considered benchmarks.

\autoref{fig:venn-compare} gives the detailed uniqueness analysis on Defects4J(v1.2) with the state-of-the-art APR approaches. 
We give the exact results for the most recent neural repair approaches (CoCoNuT, CURE and Recoder).
We combine the rest  of the top-ranked related work in a unique bin for sake of readability.
As shown,  RewardRepair fixes 5 unique bugs compared with the other APR approaches on Defects4J(v1.2). Notably, they come from three different Defects4J projects (Closure, Lang and Math) showing that the additional learned knowledge is not specific to one single domain.

\autoref{lst:Closure1} gives the RewardRepair patch for \textit{Closure-1}, which can only be fixed by RewardRepair. This is a patch with an addition of an \texttt{if} block, including the code of the \texttt{then} statement. RewardRepair learns the \texttt{if} condition from the given context code.  
While pattern-based repair is able to synthesize conditions (e.g., TBar \cite{tbar}), no pattern-based repair systems have this complete \texttt{if/then/return} pattern. 
The recent neural repair models, CoCoNuT and CURE, do not generate this patch, we suspect that with a strict token-based cross-entropy optimization, Recoder does not learn such a complex patch structure. 
\approach  is the first to produce this addition-only patch based on a non-trivial \texttt{if/then/return} structure.

\begin{table}[t!]
\small
\renewcommand{\arraystretch}{1.78}
 \begin{tabular}{p{0.35\linewidth} p{0.15\linewidth} p{0.15\linewidth} p{0.18\linewidth}}
\hline
Model &  \textbf{Top-30} &  \textbf{Top-100} &   \textbf{Top-200} \\

\hline
SequenceR~\cite{SEQUENCER} & 33\% &- &- \\
CoCoNuT~\cite{CoCoNuT} & 24\% & 15\% & 6\%-15\% \\
CURE~\cite{CURE-icse21} & 39\%  & 28\% & 14\%-28\%  \\
\hline
RewardRepair& \textbf{45.3\%}  &\textbf{37.5\%} &\textbf{33.1\%}  \\
\hline

\end{tabular}
\caption{Average compilable rates of the Top-K candidate patches in Defects4J(v1.2) and QuixBugs. ‘-’ indicates data unavailability.}	
\label{tab:compilability}
\end{table}



\textbf{Generalizability}.
Durieux et al. \cite{Durieux:2019:RepairThemAll} revealed the phenomenon of ``benchmark overfitting" in program repair, meaning that performance results reported in the APR literature do not generalize to other benchmarks. The main reason is that APR approaches were typically evaluated in a single dataset, in particular, Defects4J(v1.2) in Java evaluation. 
RewardRepair is evaluated on four benchmarks in order to maximize the generalizability of our claims. 
To our knowledge, this is one of the experiments with the largest number of testing benchmarks used for assessing the proposed repair approach.

\vspace{0.2cm}

\begin{mdframed}
Answer  to  RQ1: 
\approach  correctly fixes 45, 45, 97 and 20 bugs on the considered Java benchmarks Defects4J(v1.2), Defects4J(v2.0), Bugs.jar and QuixBugs, respectively.
There are 121 unique bugs that are repaired by RewardRepair for the
first time ever w.r.t. the APR literature.
The external validity of our results is founded on 4 testing benchmarks.
\end{mdframed}

\begin{figure}[t!]
     \includegraphics[width=0.5\textwidth]{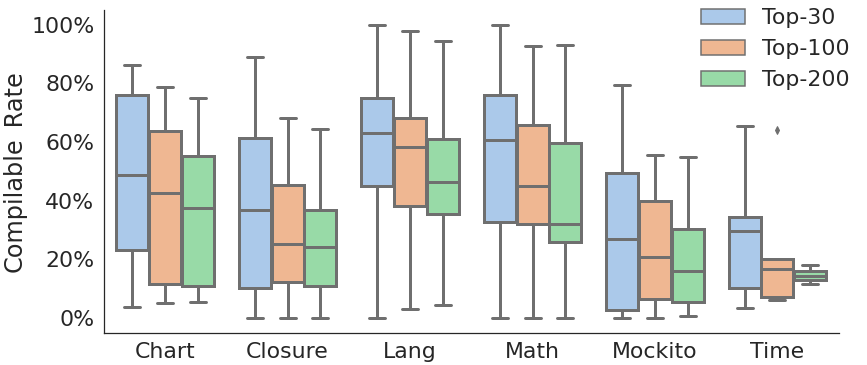}
     \caption{RewardRepair compilable rate on Defects4J(v1.2) by project.}
     \label{fig:compilable-rate}
\end{figure}

%

\vspace{1cm}

\subsection{RQ2: Improvement of Compilable Rate}

\begin{listing}[t!]
\noindent\begin{minipage}[b]{0.49\textwidth}
\begin{lstlisting} [firstnumber=419] 
 for (FormattingOption formattingOption : flags.formatting) {        
      formattingOption.applyToOptions(options); }
<@\colorbox{red!30!}{- if (flags.process\_closure\_primitives) \{ \quad\quad \quad \quad\quad\quad\quad \quad \quad\quad \quad\quad \quad\quad\quad }@>
<@\colorbox{red!30!}{-  \quad options.closurePass = true; \quad\quad \quad\quad \quad\quad \quad \quad\quad\quad\quad \quad \quad\quad \quad\quad \quad \quad\quad \quad }@>
<@\colorbox{red!30!}{-\quad \}\quad\quad \quad\quad \quad\quad \quad \quad\quad \quad\quad\quad\quad \quad\quad \quad\quad \quad \quad\quad\quad\quad \quad\quad\quad\quad\quad \quad\quad \quad \quad\quad\quad\quad}@>
<@\colorbox{green!30!}{+  options.closurePass = flags.process\_closure\_primitives; \quad\quad \quad \quad\quad\quad \quad }@>
  initOptionsFromFlags(options);
  return options;
 }
\end{lstlisting}
\end{minipage}%
\caption{\approach correct patch generated for Closure-101}
\label{lst:closure-101}
\end{listing}

\autoref{tab:compilability} shows the average compilable rates of the top-k candidate patches where k relates to the beam size. For sake of a fair comparison, we use  the same methodology as Jiang et al. \cite{CURE-icse21} and we combine Defects4J(v1.2) and QuixBugs together (our appendix website gives the results per benchmark \cite{experiment}). 
We provide ranges for CoCoNuT and CURE regarding the top-200 results due to data unavailability:
the range is the bracket [top-100,  top-1000] of compilable rates as reported in the original paper of CURE \cite{CURE-icse21}.

Notably,  the compilable rate of RewardRepair outperforms the three considered approaches CURE, CoCoNuT and SequenceR for all beam size configurations.
In the best case, RewardRepair achieves a compilable rate of up to 45.3\%, which is the highest among the three considered beam sizes. 
Next, we see that the compilable rate of \approach decreases when we increase the beam size,  this result is consistent with the ones of CURE and CoCoNuT. Since the beam enumerates by decreasing probability, it suggests that the learned neural model does capture compilability and favors it.

Recall that the key contribution of CURE \cite{CURE-icse21} is to introduce  two strategies to increase the patch compilable rate of NMT based neural repair models: \textit{valid-identifier checker strategy} and \textit{length-control strategy}.
Both strategies work in the inference phase, by filtering out the invalid tokens in the Java code or patches that are not
close in length to the buggy code.
This means that CURE does not embed program-specific knowledge in the neural network. 
On the contrary, RewardRepair learns this knowledge during semantic training.  That is if an invalid identifier is used during semantic training  and results in a non-compilable patch, RewardRepair punishes the patch by increasing the loss. 
Beyond identifiers, RewardRepair is also able to learn other programming knowledge during semantic training, such as structure and typing constraints related to the domain classes and methods.
To some extent, CURE is limited to the static analysis checks devised and implemented by its authors while RewardRepair works in a fully agnostic, data-driven way to identify important compilation constraints.

\ASERevision{We analyze the 1883 uncompilable patches from the top-30 patches generated for Defects4J(v1.2).  We show the most frequent 10 compilation errors in
\autoref{tab:uncompilable-analysis}.  If there are multiple errors for one patch, we only count the first error per patch.
The first column gives the compilation error type and the second column shows the number of patches that fail on the corresponding error type. 
We see most common error types are related to semantics that very few errors are related to syntax (the first one being ";" expected). To overcome the typing problems, this suggests doing some specific training on the project under repair to capture this missing knowledge \cite{selfapr}.}

\autoref{lst:closure-101} shows the RewardRepair patch for bug \textit{Closure-101} from Defects4J(v1.2), which is identical to the developer patch. RewardRepair incorporates two repair actions in this patch: First,  RewardRepair removes the \textit{if/then} condition. Second, RewardRepair generates a new statement by combining the logical expression from the \textit{if-condition} and from the statement in the \textit{then} block.  This is arguably a complex patch, and no repair system has reported generating a patch for this bug. This case shows that RewardRepair generates a compilable and correct patch with the complex structure. CURE fails to generate this patch because of its \textit{length-control strategy} that encourages patches similar to the buggy code in length.

\begin{table}[t!]
\footnotesize
\renewcommand{\arraystretch}{1.28}
 \begin{tabular}{p{0.75\linewidth} r }
\hline
Compile Errors &No. Failures\\
\hline
cannot find symbol&606\\
illegal start of expression&329\\
no suitable method constructor found for...&132\\
incompatible types&123\\
not a statement&102\\
";" expected&76\\
unreachable statement&63\\
case, default,or \} expected&59\\
incomparable types&56\\
method X in class Y cannot be applied to given types&53\\
\hline

\end{tabular}
\caption{Analysis of Top-10 reasons for uncompilable generated patches after semantic training.}	
\label{tab:uncompilable-analysis}
\end{table}

\begin{table}[t!]
\footnotesize
\renewcommand{\arraystretch}{2.2}
\begin{tabular}{c p{0.1\linewidth}p{0.1\linewidth}p{0.1\linewidth}p{0.1\linewidth}p{0.1\linewidth}}

\hline
\textbf{Model} & \textbf{D4J(v1.2)} &\textbf{D4J(v2.0)}&\textbf{Bugs.jar}& \textbf{QuixBugs} & \textbf{Total} \\
\hline
\makecell{RewardRepair\\(Syntatic)} & 42 &  40 & 93  &18 & 193 \\

\makecell{RewardRepair\\(Syntatic+Semantic)} & 45 &45 & 97 & 20 & 207  \\
\hline

\end{tabular}
\caption{Ablation study w.r.t correct patches. }	
\label{tab:ablation}
\end{table}


\autoref{fig:compilable-rate} shows the compilable rate of RewardRepair per project of Defects4J(v1.2) with top 30, 100 and 200 candidate patches according to beam. 
We make the two observations as follows:
First, for all  projects, increasing the beam size of RewardRepair decreases the compilable rate for each project. This confirms the conclusion made in \autoref{tab:compilability} at the level of aggregate results over bugs and benchmarks.
Second, the range of compilable rates over bugs decreases with beam size, both the range of extreme values (whiskers) and the range of interquartile values (boxes). 
We explain this by statistical sampling (sampling 30 items yields less stable results than sampling 200). However, it may also be that the compilable rate does change significantly for some bugs. 
This latter explanation is supported by the fact that there is a clear difference in compilable rate depending on the project (Lang patches compile much more than Time patches). 
This latter phenomenon -- the compilable rate significantly varying over projects, in the worst case being 0\% -- is a yet unknown limitation of neural program repair and 
suggests more future research on this to increase the compilable rate in a more uniform way.

\begin{mdframed}
Answer to RQ2: 
For all considered beam sizes, RewardRepair improves the compilable rate over the state-of-the-art.
Over all benchmarks, RewardRepair reaches up to 45.3\% of compilable patches (approximately one out of two patches compile), showing that the RewardRepair neural model has captured important information w.r.t. compilation.
\end{mdframed}

\subsection{RQ3: Impact of Semantic Training}

\autoref{tab:ablation} shows the results of the ablation study w.r.t semantic training. 
Per the same protocol as RQ1, the considered metric is the number of correct patches.
The first row shows RewardRepair's effectiveness with only syntactic training, the second row shows  RewardRepair with both syntactic and semantic training. 
For example, RewardRepair with only
syntactic training generates 42 correct patches on Defects4J(v1.2),
and RewardRepair with semantic training generates 45 correct
patches.
Overall, the addition of semantic training after syntactic training does yield more correct patches on  all considered benchmarks. This  shows that semantic training addresses the limits of syntactic training, and the improvement is not tied to specific benchmarks.

\begin{table}[t]
\footnotesize
\renewcommand{\arraystretch}{1.1}
\begin{tabular}{p{0.45\linewidth}cc}
\hline
 Actions & Syntactic Training & Semantic Training  \\
 \hline
 
\multirow{1}{*}{Add if conditions} &\multirow{1}{*}{1852} &\multirow{1}{*}{2316} \\ 

Method invocation & 1040 & 1352  \\

Add return statement & 1123 & 1554  \\


Ternary operator for null checking & 0&212\\
\hline
\end{tabular}
\caption{Examples of repair action differences between syntactic training and semantic training of RewardRepair.}
\label{tab:ablation-cases}
\end{table}

\begin{listing}[t!]
\noindent\begin{minipage}[b]{0.5\textwidth}
    \begin{lstlisting} [firstnumber=419] 
<@\colorbox{red!30}{- this.dataset = dataset; \quad\quad \quad\quad \quad\quad \quad \quad\quad \quad\quad\quad\quad\quad \quad\quad}@>           
    \end{lstlisting}
 
     \subcaption{Buggy line}
   \label{motivate-human-patch}  
    \end{minipage}%
    \hfill
    \begin{minipage}[b]{0.49\textwidth}
    \begin{lstlisting}[firstnumber=419] 
<@\colorbox{green!30!}{+  .setDataSet(dataset) \quad \quad\quad\quad\quad\quad\quad\quad\quad \quad\quad\quad\quad\quad \quad\quad \quad\quad}@> // patch 1
<@\colorbox{green!30!}{+   setDataset( ); \quad\quad \quad\quad \quad\quad \quad \quad\quad\quad\quad\quad\quad\quad\quad \quad\quad\quad\quad \quad\quad}@> // patch 2
    \end{lstlisting}
 
    \subcaption{Two non-compilable patches by syntactic training}
 \label{motivate-noncompile-patch}   
\end{minipage}%
\hfill
    \begin{minipage}[b]{0.49\textwidth}
    \begin{lstlisting}[firstnumber=419] 
<@\colorbox{green!30!}{+ setDataset(dataset); \quad\quad \quad\quad \quad\quad \quad \quad\quad \quad\quad\quad\quad \quad\quad\quad\quad\quad}@>
    \end{lstlisting} 
    \subcaption{Correct compilable patch by semantic training}
    \label{motivate-correct-patch} 
\end{minipage}%

\caption{Bug Chart-12 only fixed by semantic training}
\label{lst:chart-12-rq3}
\end{listing}

To better understand the effectiveness of semantic training, we manually analyze those  unique bugs that are only generated by RewardRepair by including semantic training (and not generated with pure syntactic training). This leads to an analysis of \numprint{14800} patches (\numprint{7400} candidate patches by syntactic training and  \numprint{7400} candidate patches by semantic training). 
We group those patches by category and summarize the most interesting categories in \autoref{tab:ablation-cases}.  
The first column gives the type of repair action employed in the patch, and the numbers in the second column and the third column indicate the numbers of patches based on those repair actions from syntactic and semantic training respectively.
For example, the first row shows there are \numprint{1852} patches generated by syntactic training which add  \textit{if/then} statements,  while semantic training yields \numprint{2316} patches using this construct. This means that the usage of \textit{if/then} statements is increased by 25.1\% with semantic training. 
Notably, the unique bugs that benefit from semantic training come from different projects,  showing that adding semantic training is beneficial in general.

Recall that syntactic training is done on more than 3 million training samples, while semantic training is done on 123 training samples, which is much less.
This suggests that the improvement obtained with semantic training does not come from the number of additional training points, but more from the training process   described in Section \ref{sec-discriminator}. RewardRepair's loss function is able to better optimize the neural network, improving the quality of generated patches, with only a few training data points.

Finally, \autoref{lst:chart-12-rq3} discusses the case of \textit{Chart-12}.
The first line is the buggy line.
Next, in part (b), two patches after syntactic training are shown, they are both close to the correct patch but none of them compile (extraneous dot in the first patch and missing parameter in the second patch). 
Finally, part (c) shows the correct patch by RewardRepair with semantic training, which is identical to the developer patch.
In this case, it suggests that the neural model with semantic training has understood that leading dots before method calls is not correct per the Java grammar, and that \texttt{setDataset} is a method likely to take a parameter called  \texttt{dataset}.
A subtle character may lead to a huge difference in program execution, but this knowledge is hard to be obtained by syntactic training with cross-entropy loss.

\begin{mdframed}
Answer to RQ3: 
Our ablation study shows semantic training of RewardRepair contributes to improving the overall effectiveness in terms of correctly fixed bugs. 
\end{mdframed}

\section{ Discussion}

\subsection{Impact of Inference Beam Size}
We investigate the impact of beam search size in the inference time, and our experiment shows a bigger beam size indeed leads to more correct patches generated, which confirms the study of Tufano et al. \cite{Tufano-tse19}.
We provide the results of configuring beam size as 500 in our online appendix repository \cite{experiment}.

\subsection{Threats to Validity}
\label{sec:threats}
A threat to external validity relates to whether the performance of \approach generalizes to arbitrary programming languages. Per the standards of the field, our approach has been tested in one language (Java) and the evaluation is carried out on established benchmarks. In principle, our approach can be applied to other programming languages and datasets. 
A threat to internal validity relates to the hyper-parameter configuration we
adopted.  To ensure replicability and extension, 
we make all the source code and results publicly available for future research ~\cite{experiment}.

\section{Related Work}

\subsection{Automatic Program Repair}

A decade of research has generated a rich body of work on automatic program repair \cite{Monperrus2015,TSE-repair-survey}. 
We have already discussed neural repair approaches  \cite{SEQUENCER,CoCoNuT,DLFix,Tufano-ICSE19,codit-tse20,Tufano-tse19,deepfix, Recoder,CURE-icse21} in Section \ref{sec:background_neural}. 
These approaches only use the syntactic cross-entropy loss objective, which poses a discrepancy between the training objective of generating compilable correct patches and the loss criterion.
The key novelty of \approach is the discriminative model to capture the compilation and execution knowledge during model training and backpropagation.

We mentioned generate and validate (G\&V) program repair approaches in RQ1 (Section \ref{sec:result_comparison}). Other notable G\&V systems include \cite{jaid,Yuan2017ARJAAR,astor,ali-issta19-bytecode}.
Moreover, the third line of research is about synthesis-based repair \cite{acs,nopol,Angelixicse16,s3,directfix,CrashProgramRepair-ISSTA19,concolic-repair-PLDI21}
which converts the search problem to a satisfiability problem.
All these approaches often work by extracting a repair constraint typically via symbolic execution incorporated with human knowledge for patch generation.
On the contrary, \approach automatically learns such fix operators,  language grammar and semantics from the training corpus.

In the field of APR, the recent work of Jiang et al. \cite{CURE-icse21} is the most closely related to ours, also focusing on the  non-compilable patch problem. They address this problem by employing a valid-identifier checker in the inference stage to filter invalid tokens.
However, many reasons could lead to a non-compilable patch, and the presence of invalid identifiers is only one of them. 
Our approach is fundamentally different: 1) RewardRepair works at training time and not at inference time; 2) RewardRepair is based on the actual compilation and test execution of training patches; 3) RewardRepair encourages syntactic diversity while CURE encourages patches similar to the buggy code (the length-control strategy in CURE).


\subsection{Discriminators for Machine Learning on Code}

Several works propose deep learning on code based on a discriminator \cite{Harer-GAN-NIPS18, adversarial-repair-gan} where the discriminator provides a loss that solves the discrepancy between the generated and real distributions of the object under study, as pioneered by generative adversarial networks (GAN)\cite{GanGoodFellow}.  
%
Harer et al. \cite{Harer-GAN-NIPS18} propose an adversarial learning approach to solve software vulnerabilities. 
Alhefdhi et al. \cite{adversarial-repair-gan} leverage a similar GAN architecture to suggest repairs that are as close as possible to human-written repairs. 
\approach shares the concepts of employing a traditional NMT model as a generator, and of replacing the cross-entropy loss with the feedback from a discriminator.  
The key difference between this related work and ours is that our discriminator uses execution information, through compilation and test execution. 

\subsection{Improving Backpropagation}
Past research has improved the cross-entropy loss based on domain-specific knowledge. 
In neural machine translation, 
Zhang et al. \cite{NMT-bridging} show the limitation of considering cross-entropy loss and its tendency to overcorrect synonymous words and phrases. To relieve the problem,  further research \cite{NMT-bridging,overcorrection} proposed to combine  cross-entropy loss and add translation evaluation at the sentence level.  
In object detection, Ryou et al. \cite{AnchorLoss} proposed AnchorLoss to dynamically rescale the cross-entropy based on prediction difficulty.
Loss scaling is a technique used in floating-point optimization, consisting of scaling up the loss value up before the start of backpropagation ~\cite{micikevicius2018mixed,lossscale}.

\subsection{Training based on Execution}

Recently, semantic information has been used in program synthesis tasks. 
Chen et al. \cite{chen2018executionguided} and Gupta et al. \cite{sed-nips20} propose execution-guided synthesis leveraging the semantics of the language.  These approaches execute a partial program to obtain intermediate states to guide program synthesis.
Wang et al. \cite{wang-sar}  use dynamic information from execution to measure semantic redundancy between student programs. 
Mesbah et al. \cite{deepdelta} extract compiler diagnostic information as an input source for repairing compilation errors. 
As in our work, these approaches use execution information as an additional input for the considered model. The key difference is that none of them employ the execution information as a reward signal to update the neural network weights through backpropagation.

Wang and colleagues \cite{wang2018dynamic,embed-ke-PLDI20} leverage the full execution traces to learn neural semantic program embeddings. These related works improve the code representation based on semantic information.
Our novelty is not on the representation, but on the training objective improvement, which is not addressed in \cite{wang2018dynamic,embed-ke-PLDI20}.


\section{Conclusion}

We have presented a novel neural program repair model \approach  based on compilation and test execution. The key idea is to employ a discriminative model to provide a reward signal on the generated patches according to the actual execution outcome. This signal modulates the purely syntactic cross-entropy loss function in what we call semantic training.  
We have conducted an extensive empirical evaluation, including a comprehensive experiment on  the widely used benchmark Defects4J, Bugs.jar and QuixBugs.  
Our results  show that it is possible to embed execution information in the backpropagation process to improve neural program repair.


\section{Acknowledgments}
\ASERevision{We thank the anonymous reviewers for the insightful feedback.
This work was supported by the Wallenberg AI, Autonomous Systems and Software Program (WASP) funded by the Knut and Alice Wallenberg Foundation.
Some experiments were performed on resources provided by the Swedish National Infrastructure for Computing.}

\bibliographystyle{ACM-Reference-Format}
\bibliography{reference}


\begin{thebibliography}{77}


\ifx \showCODEN    \undefined \def \showCODEN     #1{\unskip}     \fi
\ifx \showDOI      \undefined \def \showDOI       #1{#1}\fi
\ifx \showISBNx    \undefined \def \showISBNx     #1{\unskip}     \fi
\ifx \showISBNxiii \undefined \def \showISBNxiii  #1{\unskip}     \fi
\ifx \showISSN     \undefined \def \showISSN      #1{\unskip}     \fi
\ifx \showLCCN     \undefined \def \showLCCN      #1{\unskip}     \fi
\ifx \shownote     \undefined \def \shownote      #1{#1}          \fi
\ifx \showarticletitle \undefined \def \showarticletitle #1{#1}   \fi
\ifx \showURL      \undefined \def \showURL       {\relax}        \fi
\providecommand\bibfield[2]{#2}
\providecommand\bibinfo[2]{#2}
\providecommand\natexlab[1]{#1}
\providecommand\showeprint[2][]{arXiv:#2}

\bibitem[\protect\citeauthoryear{Abreu, Zoeteweij, and van Gemund}{Abreu
  et~al\mbox{.}}{2007}]%
        {fl-tool}
\bibfield{author}{\bibinfo{person}{Rui Abreu}, \bibinfo{person}{Peter
  Zoeteweij}, {and} \bibinfo{person}{Arjan~J.C. van Gemund}.}
  \bibinfo{year}{2007}\natexlab{}.
\newblock \showarticletitle{On the Accuracy of Spectrum-based Fault
  Localization}. In \bibinfo{booktitle}{\emph{Testing: Academic and Industrial
  Conference Practice and Research Techniques - MUTATION (TAICPART-MUTATION
  2007)}}. \bibinfo{pages}{89--98}.
\newblock


\bibitem[\protect\citeauthoryear{Alhefdhi, Dam, Le, and Ghose}{Alhefdhi
  et~al\mbox{.}}{2020}]%
        {adversarial-repair-gan}
\bibfield{author}{\bibinfo{person}{Abdulaziz Alhefdhi},
  \bibinfo{person}{Hoa~Khanh Dam}, \bibinfo{person}{Xuan-Bach~D. Le}, {and}
  \bibinfo{person}{Aditya Ghose}.} \bibinfo{year}{2020}\natexlab{}.
\newblock \bibinfo{title}{Adversarial Patch Generation for Automatic Program
  Repair}.
\newblock
\newblock
\showeprint[arxiv]{2012.11060}


\bibitem[\protect\citeauthoryear{Bahdanau, Cho, and Bengio}{Bahdanau
  et~al\mbox{.}}{2014}]%
        {nmt}
\bibfield{author}{\bibinfo{person}{Dzmitry Bahdanau},
  \bibinfo{person}{Kyunghyun Cho}, {and} \bibinfo{person}{Y. Bengio}.}
  \bibinfo{year}{2014}\natexlab{}.
\newblock \showarticletitle{Neural Machine Translation by Jointly Learning to
  Align and Translate}.
\newblock \bibinfo{journal}{\emph{ArXiv}}  \bibinfo{volume}{1409}
  (\bibinfo{date}{09} \bibinfo{year}{2014}).
\newblock


\bibitem[\protect\citeauthoryear{Baudry, Chen, Etemadi, Fu, Ginelli, Kommrusch,
  Martinez, Martin, Ron, Ye, and Yu}{Baudry et~al\mbox{.}}{2021}]%
        {Rhero}
\bibfield{author}{\bibinfo{person}{B. Baudry}, \bibinfo{person}{Zimin Chen},
  \bibinfo{person}{K. Etemadi}, \bibinfo{person}{Han Fu},
  \bibinfo{person}{Davide Ginelli}, \bibinfo{person}{Steve Kommrusch},
  \bibinfo{person}{Matias Martinez}, \bibinfo{person}{Monperrus Martin},
  \bibinfo{person}{Javier Ron}, \bibinfo{person}{He Ye}, {and}
  \bibinfo{person}{Zhongxing Yu}.} \bibinfo{year}{2021}\natexlab{}.
\newblock \showarticletitle{A Software-Repair Robot Based on Continual
  Learning}.
\newblock \bibinfo{journal}{\emph{IEEE Software}}  \bibinfo{volume}{38}
  (\bibinfo{year}{2021}), \bibinfo{pages}{28--35}.
\newblock


\bibitem[\protect\citeauthoryear{{Chakraborty}, {Ding}, {Allamanis}, and
  {Ray}}{{Chakraborty} et~al\mbox{.}}{2020}]%
        {codit-tse20}
\bibfield{author}{\bibinfo{person}{S. {Chakraborty}}, \bibinfo{person}{Y.
  {Ding}}, \bibinfo{person}{M. {Allamanis}}, {and} \bibinfo{person}{B. {Ray}}.}
  \bibinfo{year}{2020}\natexlab{}.
\newblock \showarticletitle{CODIT: Code Editing with Tree-Based Neural Models}.
\newblock \bibinfo{journal}{\emph{IEEE Transactions on Software Engineering}}
  (\bibinfo{year}{2020}).
\newblock
\urldef\tempurl%
\url{https://doi.org/10.1109/TSE.2020.3020502}
\showDOI{\tempurl}


\bibitem[\protect\citeauthoryear{Chen, Pei, and Furia}{Chen
  et~al\mbox{.}}{2017}]%
        {jaid}
\bibfield{author}{\bibinfo{person}{L. Chen}, \bibinfo{person}{Y. Pei}, {and}
  \bibinfo{person}{C.~A. Furia}.} \bibinfo{year}{2017}\natexlab{}.
\newblock \showarticletitle{Contract-based program repair without the
  contracts}. In \bibinfo{booktitle}{\emph{2017 32nd IEEE/ACM International
  Conference on Automated Software Engineering (ASE)}}.
\newblock


\bibitem[\protect\citeauthoryear{Chen, Liu, and Song}{Chen
  et~al\mbox{.}}{2019}]%
        {chen2018executionguided}
\bibfield{author}{\bibinfo{person}{Xinyun Chen}, \bibinfo{person}{Chang Liu},
  {and} \bibinfo{person}{Dawn Song}.} \bibinfo{year}{2019}\natexlab{}.
\newblock \showarticletitle{Execution-Guided Neural Program Synthesis}. In
  \bibinfo{booktitle}{\emph{International Conference on Learning
  Representations}}.
\newblock


\bibitem[\protect\citeauthoryear{{Chen}, {Kommrusch}, {Tufano}, {Pouchet},
  {Poshyvanyk}, and {Monperrus}}{{Chen} et~al\mbox{.}}{2019}]%
        {SEQUENCER}
\bibfield{author}{\bibinfo{person}{Z. {Chen}}, \bibinfo{person}{S.~J.
  {Kommrusch}}, \bibinfo{person}{M. {Tufano}}, \bibinfo{person}{L. {Pouchet}},
  \bibinfo{person}{D. {Poshyvanyk}}, {and} \bibinfo{person}{M. {Monperrus}}.}
  \bibinfo{year}{2019}\natexlab{}.
\newblock \showarticletitle{SEQUENCER: Sequence-to-Sequence Learning for
  End-to-End Program Repair}.
\newblock \bibinfo{journal}{\emph{IEEE Transactions on Software Engineering}}
  (\bibinfo{year}{2019}).
\newblock


\bibitem[\protect\citeauthoryear{Chen and Monperrus}{Chen and
  Monperrus}{2018}]%
        {Chen2018Coderep}
\bibfield{author}{\bibinfo{person}{Zimin Chen} {and} \bibinfo{person}{Martin
  Monperrus}.} \bibinfo{year}{2018}\natexlab{}.
\newblock \bibinfo{booktitle}{\emph{The CodRep Machine Learning on Source Code
  Competition}}.
\newblock \bibinfo{type}{{T}echnical {R}eport} 1807.03200.
  \bibinfo{institution}{arXiv}.
\newblock
\urldef\tempurl%
\url{http://arxiv.org/pdf/1807.03200}
\showURL{%
\tempurl}


\bibitem[\protect\citeauthoryear{Chen and Monperrus}{Chen and
  Monperrus}{2019}]%
        {machineonlearningoncodesurvey}
\bibfield{author}{\bibinfo{person}{Zimin Chen} {and} \bibinfo{person}{Martin
  Monperrus}.} \bibinfo{year}{2019}\natexlab{}.
\newblock \bibinfo{title}{A Literature Study of Embeddings on Source Code}.
\newblock
\newblock
\showeprint[arxiv]{1904.03061}


\bibitem[\protect\citeauthoryear{Durieux, Madeiral, Martinez, and
  Abreu}{Durieux et~al\mbox{.}}{2019}]%
        {Durieux:2019:RepairThemAll}
\bibfield{author}{\bibinfo{person}{Thomas Durieux}, \bibinfo{person}{Fernanda
  Madeiral}, \bibinfo{person}{Matias Martinez}, {and} \bibinfo{person}{Rui
  Abreu}.} \bibinfo{year}{2019}\natexlab{}.
\newblock \showarticletitle{Empirical Review of Java Program Repair Tools: A
  Large-scale Experiment on 2,141 Bugs and 23,551 Repair Attempts}. In
  \bibinfo{booktitle}{\emph{Proceedings of the 2019 27th ACM Joint Meeting on
  European Software Engineering Conference and Symposium on the Foundations of
  Software Engineering}}. \bibinfo{publisher}{ACM}, \bibinfo{pages}{302--313}.
\newblock
\showISBNx{978-1-4503-5572-8}
\urldef\tempurl%
\url{https://doi.org/10.1145/3338906.3338911}
\showDOI{\tempurl}


\bibitem[\protect\citeauthoryear{Experiment}{Experiment}{2022}]%
        {experiment}
\bibfield{author}{\bibinfo{person}{RewardRepair Experiment}.}
  \bibinfo{year}{2022}\natexlab{}.
\newblock \bibinfo{title}{Appendix Repository}.
\newblock
\newblock
\urldef\tempurl%
\url{https://anonymous.4open.science/r/RewardRepair}
\showURL{%
\tempurl}


\bibitem[\protect\citeauthoryear{Gao, Mechtaev, and Roychoudhury}{Gao
  et~al\mbox{.}}{2019}]%
        {CrashProgramRepair-ISSTA19}
\bibfield{author}{\bibinfo{person}{Xiang Gao}, \bibinfo{person}{Sergey
  Mechtaev}, {and} \bibinfo{person}{Abhik Roychoudhury}.}
  \bibinfo{year}{2019}\natexlab{}.
\newblock \showarticletitle{Crash-Avoiding Program Repair}. In
  \bibinfo{booktitle}{\emph{Proceedings of the 28th ACM SIGSOFT International
  Symposium on Software Testing and Analysis}} (Beijing, China)
  \emph{(\bibinfo{series}{ISSTA 2019})}. \bibinfo{publisher}{Association for
  Computing Machinery}, \bibinfo{address}{New York, NY, USA},
  \bibinfo{pages}{8–18}.
\newblock
\showISBNx{9781450362245}
\urldef\tempurl%
\url{https://doi.org/10.1145/3293882.3330558}
\showDOI{\tempurl}


\bibitem[\protect\citeauthoryear{Gazzola, Micucci, and Mariani}{Gazzola
  et~al\mbox{.}}{2017}]%
        {TSE-repair-survey}
\bibfield{author}{\bibinfo{person}{Luca Gazzola}, \bibinfo{person}{Daniela
  Micucci}, {and} \bibinfo{person}{Leonardo Mariani}.}
  \bibinfo{year}{2017}\natexlab{}.
\newblock \showarticletitle{Automatic Software Repair: A Survey}.
\newblock \bibinfo{journal}{\emph{IEEE Transactions on Software Engineering}}
  (\bibinfo{year}{2017}).
\newblock


\bibitem[\protect\citeauthoryear{Ghanbari, Benton, and Zhang}{Ghanbari
  et~al\mbox{.}}{2019}]%
        {ali-issta19-bytecode}
\bibfield{author}{\bibinfo{person}{Ali Ghanbari}, \bibinfo{person}{Samuel
  Benton}, {and} \bibinfo{person}{Lingming Zhang}.}
  \bibinfo{year}{2019}\natexlab{}.
\newblock \showarticletitle{Practical Program Repair via Bytecode Mutation}. In
  \bibinfo{booktitle}{\emph{Proceedings of the 28th ACM SIGSOFT International
  Symposium on Software Testing and Analysis}} (Beijing, China)
  \emph{(\bibinfo{series}{ISSTA 2019})}. \bibinfo{publisher}{Association for
  Computing Machinery}, \bibinfo{address}{New York, NY, USA},
  \bibinfo{pages}{19–30}.
\newblock
\showISBNx{9781450362245}
\urldef\tempurl%
\url{https://doi.org/10.1145/3293882.3330559}
\showDOI{\tempurl}


\bibitem[\protect\citeauthoryear{Ghazvininejad, Karpukhin, Zettlemoyer, and
  Levy}{Ghazvininejad et~al\mbox{.}}{2020}]%
        {aligncrossentropy}
\bibfield{author}{\bibinfo{person}{Marjan Ghazvininejad},
  \bibinfo{person}{Vladimir Karpukhin}, \bibinfo{person}{Luke Zettlemoyer},
  {and} \bibinfo{person}{Omer Levy}.} \bibinfo{year}{2020}\natexlab{}.
\newblock \bibinfo{title}{Aligned Cross Entropy for Non-Autoregressive Machine
  Translation}.
\newblock
\newblock
\showeprint[arxiv]{2004.01655}~[cs.CL]


\bibitem[\protect\citeauthoryear{Goodfellow, Pouget-Abadie, Mirza, Xu,
  Warde-Farley, Ozair, Courville, and Bengio}{Goodfellow et~al\mbox{.}}{2014}]%
        {GanGoodFellow}
\bibfield{author}{\bibinfo{person}{Ian~J. Goodfellow}, \bibinfo{person}{Jean
  Pouget-Abadie}, \bibinfo{person}{Mehdi Mirza}, \bibinfo{person}{Bing Xu},
  \bibinfo{person}{David Warde-Farley}, \bibinfo{person}{Sherjil Ozair},
  \bibinfo{person}{Aaron Courville}, {and} \bibinfo{person}{Yoshua Bengio}.}
  \bibinfo{year}{2014}\natexlab{}.
\newblock \showarticletitle{Generative Adversarial Nets}
  \emph{(\bibinfo{series}{NIPS'14})}. \bibinfo{publisher}{MIT Press},
  \bibinfo{address}{Cambridge, MA, USA}, \bibinfo{pages}{2672–2680}.
\newblock


\bibitem[\protect\citeauthoryear{Gupta, Christensen, Chen, and Song}{Gupta
  et~al\mbox{.}}{2020}]%
        {sed-nips20}
\bibfield{author}{\bibinfo{person}{Kavi Gupta}, \bibinfo{person}{Peter~Ebert
  Christensen}, \bibinfo{person}{Xinyun Chen}, {and} \bibinfo{person}{Dawn
  Song}.} \bibinfo{year}{2020}\natexlab{}.
\newblock \showarticletitle{Synthesize, Execute and Debug: Learning to Repair
  for Neural Program Synthesis} \emph{(\bibinfo{series}{NIPS'20})}.
\newblock


\bibitem[\protect\citeauthoryear{Gupta, Pal, Kanade, and Shevade}{Gupta
  et~al\mbox{.}}{2017}]%
        {deepfix}
\bibfield{author}{\bibinfo{person}{Rahul Gupta}, \bibinfo{person}{Soham Pal},
  \bibinfo{person}{Aditya Kanade}, {and} \bibinfo{person}{Shirish Shevade}.}
  \bibinfo{year}{2017}\natexlab{}.
\newblock \showarticletitle{DeepFix: Fixing Common C Language Errors by Deep
  Learning}. In \bibinfo{booktitle}{\emph{Proceedings of the Thirty-First AAAI
  Conference on Artificial Intelligence}} (San Francisco, California, USA)
  \emph{(\bibinfo{series}{AAAI'17})}. \bibinfo{publisher}{AAAI Press},
  \bibinfo{pages}{1345–1351}.
\newblock


\bibitem[\protect\citeauthoryear{Harer, Ozdemir, Lazovich, Reale, Russell, Kim,
  and Chin}{Harer et~al\mbox{.}}{2018}]%
        {Harer-GAN-NIPS18}
\bibfield{author}{\bibinfo{person}{Jacob~A. Harer}, \bibinfo{person}{Onur
  Ozdemir}, \bibinfo{person}{Tomo Lazovich}, \bibinfo{person}{Christopher~P.
  Reale}, \bibinfo{person}{Rebecca~L. Russell}, \bibinfo{person}{Louis~Y. Kim},
  {and} \bibinfo{person}{Peter Chin}.} \bibinfo{year}{2018}\natexlab{}.
\newblock \showarticletitle{Learning to Repair Software Vulnerabilities with
  Generative Adversarial Networks} \emph{(\bibinfo{series}{NIPS'18})}.
  \bibinfo{publisher}{Curran Associates Inc.}, \bibinfo{address}{Red Hook, NY,
  USA}, \bibinfo{pages}{7944–7954}.
\newblock


\bibitem[\protect\citeauthoryear{Jiang, Xiong, Zhang, Gao, and Chen}{Jiang
  et~al\mbox{.}}{2018}]%
        {Simfix:2018}
\bibfield{author}{\bibinfo{person}{Jiajun Jiang}, \bibinfo{person}{Yingfei
  Xiong}, \bibinfo{person}{Hongyu Zhang}, \bibinfo{person}{Qing Gao}, {and}
  \bibinfo{person}{Xiangqun Chen}.} \bibinfo{year}{2018}\natexlab{}.
\newblock \showarticletitle{Shaping Program Repair Space with Existing Patches
  and Similar Code} \emph{(\bibinfo{series}{ISSTA})}.
\newblock


\bibitem[\protect\citeauthoryear{Jiang, Lutellier, and Tan}{Jiang
  et~al\mbox{.}}{2021}]%
        {CURE-icse21}
\bibfield{author}{\bibinfo{person}{Nan Jiang}, \bibinfo{person}{Thibaud
  Lutellier}, {and} \bibinfo{person}{Lin Tan}.}
  \bibinfo{year}{2021}\natexlab{}.
\newblock \showarticletitle{CURE: Code-Aware Neural Machine Translation for
  Automatic Program Repair}. In \bibinfo{booktitle}{\emph{Proceedings of the
  ACM/IEEE 43rd International Conference on Software Engineering}}.
\newblock


\bibitem[\protect\citeauthoryear{Just, Jalali, and Ernst}{Just
  et~al\mbox{.}}{2014}]%
        {defects4j}
\bibfield{author}{\bibinfo{person}{Rene Just}, \bibinfo{person}{Darioush
  Jalali}, {and} \bibinfo{person}{Michael~D Ernst}.}
  \bibinfo{year}{2014}\natexlab{}.
\newblock \showarticletitle{Defects4J: A database of existing faults to enable
  controlled testing studies for Java programs}. In
  \bibinfo{booktitle}{\emph{Proceedings of the 2014 International Symposium on
  Software Testing and Analysis}}. ACM, \bibinfo{pages}{437--440}.
\newblock


\bibitem[\protect\citeauthoryear{Karampatsis, Babii, Robbes, Sutton, and
  Janes}{Karampatsis et~al\mbox{.}}{2020}]%
        {karampatsis2020big}
\bibfield{author}{\bibinfo{person}{Rafael-Michael Karampatsis},
  \bibinfo{person}{Hlib Babii}, \bibinfo{person}{Romain Robbes},
  \bibinfo{person}{Charles Sutton}, {and} \bibinfo{person}{Andrea Janes}.}
  \bibinfo{year}{2020}\natexlab{}.
\newblock \showarticletitle{Big code!= big vocabulary: Open-vocabulary models
  for source code}. In \bibinfo{booktitle}{\emph{2020 IEEE/ACM 42nd
  International Conference on Software Engineering (ICSE)}}. IEEE,
  \bibinfo{pages}{1073--1085}.
\newblock


\bibitem[\protect\citeauthoryear{Kommrusch, Monperrus, and Pouchet}{Kommrusch
  et~al\mbox{.}}{2021}]%
        {kommrusch2021selfsupervised}
\bibfield{author}{\bibinfo{person}{Steve Kommrusch}, \bibinfo{person}{Martin
  Monperrus}, {and} \bibinfo{person}{Louis-Noël Pouchet}.}
  \bibinfo{year}{2021}\natexlab{}.
\newblock \bibinfo{title}{Self-Supervised Learning to Prove Equivalence Between
  Programs via Semantics-Preserving Rewrite Rules}.
\newblock
\newblock
\showeprint[arxiv]{2109.10476}~[cs.LG]


\bibitem[\protect\citeauthoryear{Kudo and Richardson}{Kudo and
  Richardson}{2018}]%
        {Kudo2018SentencePieceAS}
\bibfield{author}{\bibinfo{person}{Taku Kudo} {and} \bibinfo{person}{John
  Richardson}.} \bibinfo{year}{2018}\natexlab{}.
\newblock \showarticletitle{SentencePiece: A simple and language independent
  subword tokenizer and detokenizer for Neural Text Processing}. In
  \bibinfo{booktitle}{\emph{Empirical Methods in Natural Language
  Processing(EMNLP)}}.
\newblock


\bibitem[\protect\citeauthoryear{Le, Chu, Lo, Le~Goues, and Visser}{Le
  et~al\mbox{.}}{2017}]%
        {s3}
\bibfield{author}{\bibinfo{person}{Xuan-Bach~D. Le}, \bibinfo{person}{Duc-Hiep
  Chu}, \bibinfo{person}{David Lo}, \bibinfo{person}{Claire Le~Goues}, {and}
  \bibinfo{person}{Willem Visser}.} \bibinfo{year}{2017}\natexlab{}.
\newblock \showarticletitle{S3: Syntax- and Semantic-guided Repair Synthesis
  via Programming by Examples}. In \bibinfo{booktitle}{\emph{Proceedings of the
  2017 11th Joint Meeting on Foundations of Software Engineering}}
  \emph{(\bibinfo{series}{ESEC/FSE 2017})}.
\newblock


\bibitem[\protect\citeauthoryear{Le, Thung, Lo, and Goues}{Le
  et~al\mbox{.}}{2018}]%
        {Le:overfitting}
\bibfield{author}{\bibinfo{person}{Xuan-Bach~D. Le}, \bibinfo{person}{Ferdian
  Thung}, \bibinfo{person}{David Lo}, {and} \bibinfo{person}{Claire~Le Goues}.}
  \bibinfo{year}{2018}\natexlab{}.
\newblock \showarticletitle{Overfitting in Semantics-based Automated Program
  Repair}. In \bibinfo{booktitle}{\emph{Proceedings of the 40th International
  Conference on Software Engineering}} (Gothenburg, Sweden)
  \emph{(\bibinfo{series}{ICSE '18})}. \bibinfo{publisher}{ACM},
  \bibinfo{address}{New York, NY, USA}.
\newblock


\bibitem[\protect\citeauthoryear{Le, Bao, Lo, Xia, and Li}{Le
  et~al\mbox{.}}{2019}]%
        {le:reliability-patch-assess}
\bibfield{author}{\bibinfo{person}{Xuan{-}Bach~D. Le},
  \bibinfo{person}{Lingfeng Bao}, \bibinfo{person}{David Lo},
  \bibinfo{person}{Xin Xia}, {and} \bibinfo{person}{Shanping Li}.}
  \bibinfo{year}{2019}\natexlab{}.
\newblock \showarticletitle{On Reliability of Patch Correctness Assessment}. In
  \bibinfo{booktitle}{\emph{Proceedings of the 41st ACM/IEEE International
  Conference on Software Engineering}}.
\newblock


\bibitem[\protect\citeauthoryear{Le~Goues, Nguyen, Forrest, and
  Weimer}{Le~Goues et~al\mbox{.}}{2012}]%
        {LeGoues2012GenProg}
\bibfield{author}{\bibinfo{person}{Claire Le~Goues}, \bibinfo{person}{ThanhVu
  Nguyen}, \bibinfo{person}{Stephanie Forrest}, {and} \bibinfo{person}{Westley
  Weimer}.} \bibinfo{year}{2012}\natexlab{}.
\newblock \showarticletitle{GenProg: A generic method for automatic software
  repair}.
\newblock \bibinfo{journal}{\emph{Software Engineering, IEEE Transactions on}}
  \bibinfo{volume}{38}, \bibinfo{number}{1} (\bibinfo{year}{2012}),
  \bibinfo{pages}{54--72}.
\newblock
\urldef\tempurl%
\url{https://doi.org/10.1109/TSE.2011.104}
\showDOI{\tempurl}


\bibitem[\protect\citeauthoryear{Li, Wang, and Nguyen}{Li
  et~al\mbox{.}}{2020}]%
        {DLFix}
\bibfield{author}{\bibinfo{person}{Yi Li}, \bibinfo{person}{Shaohua Wang},
  {and} \bibinfo{person}{Tien~N. Nguyen}.} \bibinfo{year}{2020}\natexlab{}.
\newblock \showarticletitle{DLFix: Context-Based Code Transformation Learning
  for Automated Program Repair}. In \bibinfo{booktitle}{\emph{Proceedings of
  the ACM/IEEE 42nd International Conference on Software Engineering}} (Seoul,
  South Korea) \emph{(\bibinfo{series}{ICSE '20})}.
  \bibinfo{publisher}{Association for Computing Machinery},
  \bibinfo{address}{New York, NY, USA}, \bibinfo{pages}{602–614}.
\newblock
\showISBNx{9781450371216}
\urldef\tempurl%
\url{https://doi.org/10.1145/3377811.3380345}
\showDOI{\tempurl}


\bibitem[\protect\citeauthoryear{Lin, Koppel, Chen, and Solar-Lezama}{Lin
  et~al\mbox{.}}{2017}]%
        {lin2017quixbugs}
\bibfield{author}{\bibinfo{person}{Derrick Lin}, \bibinfo{person}{James
  Koppel}, \bibinfo{person}{Angela Chen}, {and} \bibinfo{person}{Armando
  Solar-Lezama}.} \bibinfo{year}{2017}\natexlab{}.
\newblock \showarticletitle{QuixBugs: A Multi-Lingual Program Repair Benchmark
  Set Based on the Quixey Challenge}. In \bibinfo{booktitle}{\emph{Proceedings
  Companion of the 2017 ACM SIGPLAN International Conference on Systems,
  Programming, Languages, and Applications: Software for Humanity}} (Vancouver,
  BC, Canada) \emph{(\bibinfo{series}{SPLASH Companion 2017})}.
  \bibinfo{publisher}{Association for Computing Machinery},
  \bibinfo{address}{New York, NY, USA}, \bibinfo{pages}{55–56}.
\newblock
\showISBNx{9781450355148}
\urldef\tempurl%
\url{https://doi.org/10.1145/3135932.3135941}
\showDOI{\tempurl}


\bibitem[\protect\citeauthoryear{Liu, Koyuncu, Kim, and Bissyand{\'e}}{Liu
  et~al\mbox{.}}{2019}]%
        {tbar}
\bibfield{author}{\bibinfo{person}{Kui Liu}, \bibinfo{person}{Anil Koyuncu},
  \bibinfo{person}{Dongsun Kim}, {and} \bibinfo{person}{Tegawend{\'e}~F.
  Bissyand{\'e}}.} \bibinfo{year}{2019}\natexlab{}.
\newblock \showarticletitle{{TBar}: Revisiting Template-based Automated Program
  Repair}. In \bibinfo{booktitle}{\emph{Proceedings of the 28th ACM SIGSOFT
  International Symposium on Software Testing and Analysis}}. ACM,
  \bibinfo{pages}{31--42}.
\newblock
\urldef\tempurl%
\url{https://doi.org/10.1145/3293882.3330577}
\showDOI{\tempurl}


\bibitem[\protect\citeauthoryear{Liu, Wang, Koyuncu, Kim, Bissyand\'{e}, Kim,
  Wu, Klein, Mao, and Traon}{Liu et~al\mbox{.}}{2020}]%
        {Liu2020Efficiency}
\bibfield{author}{\bibinfo{person}{Kui Liu}, \bibinfo{person}{Shangwen Wang},
  \bibinfo{person}{Anil Koyuncu}, \bibinfo{person}{Kisub Kim},
  \bibinfo{person}{Tegawend\'{e}~F. Bissyand\'{e}}, \bibinfo{person}{Dongsun
  Kim}, \bibinfo{person}{Peng Wu}, \bibinfo{person}{Jacques Klein},
  \bibinfo{person}{Xiaoguang Mao}, {and} \bibinfo{person}{Yves~Le Traon}.}
  \bibinfo{year}{2020}\natexlab{}.
\newblock \showarticletitle{On the Efficiency of Test Suite Based Program
  Repair: A Systematic Assessment of 16 Automated Repair Systems for Java
  Programs}. In \bibinfo{booktitle}{\emph{Proceedings of the ACM/IEEE 42nd
  International Conference on Software Engineering}} (Seoul, South Korea)
  \emph{(\bibinfo{series}{ICSE '20})}. \bibinfo{publisher}{Association for
  Computing Machinery}, \bibinfo{address}{New York, NY, USA},
  \bibinfo{pages}{615–627}.
\newblock
\showISBNx{9781450371216}
\urldef\tempurl%
\url{https://doi.org/10.1145/3377811.3380338}
\showDOI{\tempurl}


\bibitem[\protect\citeauthoryear{Lu, Zhou, Liu, and Zhang}{Lu
  et~al\mbox{.}}{2020}]%
        {overcorrection}
\bibfield{author}{\bibinfo{person}{Wenjie Lu}, \bibinfo{person}{Leiying Zhou},
  \bibinfo{person}{Gongshen Liu}, {and} \bibinfo{person}{Quanhai Zhang}.}
  \bibinfo{year}{2020}\natexlab{}.
\newblock \showarticletitle{A Mixed Learning Objective for Neural Machine
  Translation}. In \bibinfo{booktitle}{\emph{Chinese Computational
  Linguistics}}, \bibfield{editor}{\bibinfo{person}{Maosong Sun},
  \bibinfo{person}{Sujian Li}, \bibinfo{person}{Yue Zhang},
  \bibinfo{person}{Yang Liu}, \bibinfo{person}{Shizhu He}, {and}
  \bibinfo{person}{Gaoqi Rao}} (Eds.). \bibinfo{publisher}{Springer
  International Publishing}, \bibinfo{address}{Cham},
  \bibinfo{pages}{201--213}.
\newblock
\showISBNx{978-3-030-63031-7}


\bibitem[\protect\citeauthoryear{Lutellier, Pham, Pang, Li, Wei, and
  Tan}{Lutellier et~al\mbox{.}}{2020}]%
        {CoCoNuT}
\bibfield{author}{\bibinfo{person}{Thibaud Lutellier},
  \bibinfo{person}{Hung~Viet Pham}, \bibinfo{person}{Lawrence Pang},
  \bibinfo{person}{Yitong Li}, \bibinfo{person}{Moshi Wei}, {and}
  \bibinfo{person}{Lin Tan}.} \bibinfo{year}{2020}\natexlab{}.
\newblock \showarticletitle{CoCoNuT: Combining Context-Aware Neural Translation
  Models Using Ensemble for Program Repair} \emph{(\bibinfo{series}{ISSTA
  2020})}.
\newblock
\showISBNx{9781450380089}


\bibitem[\protect\citeauthoryear{Madeiral, Urli, Maia, and Monperrus}{Madeiral
  et~al\mbox{.}}{2019}]%
        {Bears}
\bibfield{author}{\bibinfo{person}{Fernanda Madeiral}, \bibinfo{person}{Simon
  Urli}, \bibinfo{person}{Marcelo Maia}, {and} \bibinfo{person}{Martin
  Monperrus}.} \bibinfo{year}{2019}\natexlab{}.
\newblock \showarticletitle{{Bears: An Extensible Java Bug Benchmark for
  Automatic Program Repair Studies}}. In \bibinfo{booktitle}{\emph{Proceedings
  of the 26th IEEE International Conference on Software Analysis, Evolution and
  Reengineering (SANER '19)}}.
\newblock
\urldef\tempurl%
\url{https://arxiv.org/abs/1901.06024}
\showURL{%
\tempurl}


\bibitem[\protect\citeauthoryear{Martinez, Durieux, Sommerard, Xuan, and
  Monperrus}{Martinez et~al\mbox{.}}{2017}]%
        {Martinez2017experiment}
\bibfield{author}{\bibinfo{person}{Matias Martinez}, \bibinfo{person}{Thomas
  Durieux}, \bibinfo{person}{Romain Sommerard}, \bibinfo{person}{Jifeng Xuan},
  {and} \bibinfo{person}{Martin Monperrus}.} \bibinfo{year}{2017}\natexlab{}.
\newblock \showarticletitle{{Automatic Repair of Real Bugs in Java: A
  Large-scale Experiment on the Defects4J Dataset}}.
\newblock \bibinfo{journal}{\emph{Empirical Software Engineering}}
  \bibinfo{volume}{22}, \bibinfo{number}{4} (\bibinfo{year}{2017}),
  \bibinfo{pages}{1936--1964}.
\newblock
\showISSN{1382-3256}
\urldef\tempurl%
\url{https://doi.org/10.1007/s10664-016-9470-4}
\showDOI{\tempurl}


\bibitem[\protect\citeauthoryear{Martinez and Monperrus}{Martinez and
  Monperrus}{2016}]%
        {astor}
\bibfield{author}{\bibinfo{person}{Matias Martinez} {and}
  \bibinfo{person}{Martin Monperrus}.} \bibinfo{year}{2016}\natexlab{}.
\newblock \showarticletitle{ASTOR: A Program Repair Library for Java}. In
  \bibinfo{booktitle}{\emph{Proceedings of ISSTA}}.
\newblock


\bibitem[\protect\citeauthoryear{{Mechtaev}, {Yi}, and
  {Roychoudhury}}{{Mechtaev} et~al\mbox{.}}{2015}]%
        {directfix}
\bibfield{author}{\bibinfo{person}{S. {Mechtaev}}, \bibinfo{person}{J. {Yi}},
  {and} \bibinfo{person}{A. {Roychoudhury}}.} \bibinfo{year}{2015}\natexlab{}.
\newblock \showarticletitle{DirectFix: Looking for Simple Program Repairs}. In
  \bibinfo{booktitle}{\emph{2015 IEEE/ACM 37th IEEE International Conference on
  Software Engineering}}, Vol.~\bibinfo{volume}{1}. \bibinfo{pages}{448--458}.
\newblock
\urldef\tempurl%
\url{https://doi.org/10.1109/ICSE.2015.63}
\showDOI{\tempurl}


\bibitem[\protect\citeauthoryear{Mechtaev, Yi, and Roychoudhury}{Mechtaev
  et~al\mbox{.}}{2016}]%
        {Angelixicse16}
\bibfield{author}{\bibinfo{person}{Sergey Mechtaev}, \bibinfo{person}{Jooyong
  Yi}, {and} \bibinfo{person}{Abhik Roychoudhury}.}
  \bibinfo{year}{2016}\natexlab{}.
\newblock \showarticletitle{Angelix: Scalable Multiline Program Patch Synthesis
  via Symbolic Analysis}. In \bibinfo{booktitle}{\emph{2016 IEEE/ACM 38th
  International Conference on Software Engineering (ICSE)}}.
\newblock


\bibitem[\protect\citeauthoryear{Mesbah, Rice, Johnston, Glorioso, and
  Aftandilian}{Mesbah et~al\mbox{.}}{2019}]%
        {deepdelta}
\bibfield{author}{\bibinfo{person}{Ali Mesbah}, \bibinfo{person}{Andrew Rice},
  \bibinfo{person}{Emily Johnston}, \bibinfo{person}{Nick Glorioso}, {and}
  \bibinfo{person}{Edward Aftandilian}.} \bibinfo{year}{2019}\natexlab{}.
\newblock \showarticletitle{DeepDelta: Learning to Repair Compilation Errors}.
  In \bibinfo{booktitle}{\emph{Proceedings of the 2019 27th ACM Joint Meeting
  on European Software Engineering Conference and Symposium on the Foundations
  of Software Engineering}} (Tallinn, Estonia) \emph{(\bibinfo{series}{ESEC/FSE
  2019})}. \bibinfo{publisher}{Association for Computing Machinery},
  \bibinfo{address}{New York, NY, USA}, \bibinfo{pages}{925–936}.
\newblock
\showISBNx{9781450355728}
\urldef\tempurl%
\url{https://doi.org/10.1145/3338906.3340455}
\showDOI{\tempurl}


\bibitem[\protect\citeauthoryear{Micikevicius, Narang, Alben, Diamos, Elsen,
  Garcia, Ginsburg, Houston, Kuchaiev, Venkatesh, and Wu}{Micikevicius
  et~al\mbox{.}}{2018}]%
        {micikevicius2018mixed}
\bibfield{author}{\bibinfo{person}{Paulius Micikevicius},
  \bibinfo{person}{Sharan Narang}, \bibinfo{person}{Jonah Alben},
  \bibinfo{person}{Gregory Diamos}, \bibinfo{person}{Erich Elsen},
  \bibinfo{person}{David Garcia}, \bibinfo{person}{Boris Ginsburg},
  \bibinfo{person}{Michael Houston}, \bibinfo{person}{Oleksii Kuchaiev},
  \bibinfo{person}{Ganesh Venkatesh}, {and} \bibinfo{person}{Hao Wu}.}
  \bibinfo{year}{2018}\natexlab{}.
\newblock \showarticletitle{Mixed Precision Training}. In
  \bibinfo{booktitle}{\emph{International Conference on Learning
  Representations}}.
\newblock
\urldef\tempurl%
\url{https://openreview.net/forum?id=r1gs9JgRZ}
\showURL{%
\tempurl}


\bibitem[\protect\citeauthoryear{Monperrus}{Monperrus}{2017}]%
        {Monperrus2015}
\bibfield{author}{\bibinfo{person}{Martin Monperrus}.}
  \bibinfo{year}{2017}\natexlab{}.
\newblock \showarticletitle{Automatic Software Repair: a Bibliography}.
\newblock \bibinfo{journal}{\emph{{ACM Computing Surveys}}}
  \bibinfo{volume}{51} (\bibinfo{year}{2017}), \bibinfo{pages}{1--24}.
\newblock
\urldef\tempurl%
\url{https://doi.org/10.1145/3105906}
\showDOI{\tempurl}


\bibitem[\protect\citeauthoryear{Monperrus, Martinez, Ye, Madeiral, Durieux,
  and Yu}{Monperrus et~al\mbox{.}}{2021}]%
        {monperrus2021megadiff}
\bibfield{author}{\bibinfo{person}{Martin Monperrus}, \bibinfo{person}{Matias
  Martinez}, \bibinfo{person}{He Ye}, \bibinfo{person}{Fernanda Madeiral},
  \bibinfo{person}{Thomas Durieux}, {and} \bibinfo{person}{Zhongxing Yu}.}
  \bibinfo{year}{2021}\natexlab{}.
\newblock \bibinfo{title}{Megadiff: A Dataset of 600k Java Source Code Changes
  Categorized by Diff Size}.
\newblock
\newblock
\showeprint[arxiv]{2108.04631}~[cs.SE]


\bibitem[\protect\citeauthoryear{Rabin, Bui, Wang, Yu, Jiang, and
  Alipour}{Rabin et~al\mbox{.}}{2021}]%
        {rabin2021generalizability}
\bibfield{author}{\bibinfo{person}{Md~Rafiqul~Islam Rabin},
  \bibinfo{person}{Nghi~D.Q. Bui}, \bibinfo{person}{Ke Wang},
  \bibinfo{person}{Yijun Yu}, \bibinfo{person}{Lingxiao Jiang}, {and}
  \bibinfo{person}{Mohammad~Amin Alipour}.} \bibinfo{year}{2021}\natexlab{}.
\newblock \showarticletitle{On the generalizability of Neural Program Models
  with respect to semantic-preserving program transformations}.
\newblock \bibinfo{journal}{\emph{Information and Software Technology}}
  \bibinfo{volume}{135} (\bibinfo{year}{2021}), \bibinfo{pages}{106552}.
\newblock
\showISSN{0950-5849}


\bibitem[\protect\citeauthoryear{Raffel, Shazeer, Roberts, Lee, Narang, Matena,
  Zhou, Li, and Liu}{Raffel et~al\mbox{.}}{2020}]%
        {T5}
\bibfield{author}{\bibinfo{person}{Colin Raffel}, \bibinfo{person}{Noam
  Shazeer}, \bibinfo{person}{Adam Roberts}, \bibinfo{person}{Katherine Lee},
  \bibinfo{person}{Sharan Narang}, \bibinfo{person}{Michael Matena},
  \bibinfo{person}{Yanqi Zhou}, \bibinfo{person}{Wei Li}, {and}
  \bibinfo{person}{Peter~J. Liu}.} \bibinfo{year}{2020}\natexlab{}.
\newblock \showarticletitle{Exploring the Limits of Transfer Learning with a
  Unified Text-to-Text Transformer}.
\newblock \bibinfo{journal}{\emph{Journal of Machine Learning Research}}
  \bibinfo{volume}{21}, \bibinfo{number}{140} (\bibinfo{year}{2020}),
  \bibinfo{pages}{1--67}.
\newblock
\urldef\tempurl%
\url{http://jmlr.org/papers/v21/20-074.html}
\showURL{%
\tempurl}


\bibitem[\protect\citeauthoryear{Riboira and Abreu}{Riboira and Abreu}{2010}]%
        {GZoltar}
\bibfield{author}{\bibinfo{person}{Andr\'{e} Riboira} {and}
  \bibinfo{person}{Rui Abreu}.} \bibinfo{year}{2010}\natexlab{}.
\newblock \showarticletitle{The GZoltar Project: A Graphical Debugger
  Interface} \emph{(\bibinfo{series}{TAIC PART'10})}.
  \bibinfo{publisher}{Springer-Verlag}, \bibinfo{address}{Berlin, Heidelberg}.
\newblock
\showISBNx{3642155847}


\bibitem[\protect\citeauthoryear{Ryou, Jeong, and Perona}{Ryou
  et~al\mbox{.}}{2019}]%
        {AnchorLoss}
\bibfield{author}{\bibinfo{person}{Serim Ryou}, \bibinfo{person}{Seong-Gyun
  Jeong}, {and} \bibinfo{person}{P. Perona}.} \bibinfo{year}{2019}\natexlab{}.
\newblock \showarticletitle{Anchor Loss: Modulating Loss Scale Based on
  Prediction Difficulty}.
\newblock \bibinfo{journal}{\emph{2019 IEEE/CVF International Conference on
  Computer Vision (ICCV)}} (\bibinfo{year}{2019}), \bibinfo{pages}{5991--6000}.
\newblock


\bibitem[\protect\citeauthoryear{Saha, Lyu, Lam, Yoshida, and Prasad}{Saha
  et~al\mbox{.}}{2018}]%
        {Bugsjar-MSR18}
\bibfield{author}{\bibinfo{person}{Ripon~K. Saha}, \bibinfo{person}{Yingjun
  Lyu}, \bibinfo{person}{Wing Lam}, \bibinfo{person}{Hiroaki Yoshida}, {and}
  \bibinfo{person}{Mukul~R. Prasad}.} \bibinfo{year}{2018}\natexlab{}.
\newblock \showarticletitle{Bugs.Jar: A Large-Scale, Diverse Dataset of
  Real-World Java Bugs} \emph{(\bibinfo{series}{MSR '18})}.
  \bibinfo{publisher}{Association for Computing Machinery},
  \bibinfo{address}{New York, NY, USA}, \bibinfo{pages}{10–13}.
\newblock
\showISBNx{9781450357166}
\urldef\tempurl%
\url{https://doi.org/10.1145/3196398.3196473}
\showDOI{\tempurl}


\bibitem[\protect\citeauthoryear{Saha, Lyu, Yoshida, and Prasad}{Saha
  et~al\mbox{.}}{2017}]%
        {elixir}
\bibfield{author}{\bibinfo{person}{Ripon~K. Saha}, \bibinfo{person}{Yingjun
  Lyu}, \bibinfo{person}{Hiroaki Yoshida}, {and} \bibinfo{person}{Mukul~R.
  Prasad}.} \bibinfo{year}{2017}\natexlab{}.
\newblock \showarticletitle{ELIXIR: Effective Object Oriented Program Repair}.
  In \bibinfo{booktitle}{\emph{Proceedings of the 32Nd IEEE/ACM International
  Conference on Automated Software Engineering}} (Urbana-Champaign, IL, USA)
  \emph{(\bibinfo{series}{ASE 2017})}. \bibinfo{publisher}{IEEE Press},
  \bibinfo{address}{Piscataway, NJ, USA}, \bibinfo{pages}{648--659}.
\newblock
\showISBNx{978-1-5386-2684-9}


\bibitem[\protect\citeauthoryear{Saha, Saha, and Prasad}{Saha
  et~al\mbox{.}}{2019}]%
        {hercules}
\bibfield{author}{\bibinfo{person}{Seemanta Saha}, \bibinfo{person}{Ripon~K.
  Saha}, {and} \bibinfo{person}{Mukul~R. Prasad}.}
  \bibinfo{year}{2019}\natexlab{}.
\newblock \showarticletitle{Harnessing Evolution for Multi-Hunk Program
  Repair}. In \bibinfo{booktitle}{\emph{Proceedings of the 41st International
  Conference on Software Engineering}} (Montreal, Quebec, Canada)
  \emph{(\bibinfo{series}{ICSE '19})}. \bibinfo{publisher}{IEEE Press},
  \bibinfo{pages}{13–24}.
\newblock
\urldef\tempurl%
\url{https://doi.org/10.1109/ICSE.2019.00020}
\showDOI{\tempurl}


\bibitem[\protect\citeauthoryear{Shariffdeen, Noller, Grunske, and
  Roychoudhury}{Shariffdeen et~al\mbox{.}}{2021}]%
        {concolic-repair-PLDI21}
\bibfield{author}{\bibinfo{person}{Ridwan Shariffdeen}, \bibinfo{person}{Yannic
  Noller}, \bibinfo{person}{Lars Grunske}, {and} \bibinfo{person}{Abhik
  Roychoudhury}.} \bibinfo{year}{2021}\natexlab{}.
\newblock \showarticletitle{Concolic Program Repair}. In
  \bibinfo{booktitle}{\emph{42nd ACM SIGPLAN Conference on Programming Language
  Design and Implementation (PLDI)}}.
\newblock


\bibitem[\protect\citeauthoryear{Smith, Barr, Le~Goues, and Brun}{Smith
  et~al\mbox{.}}{2015}]%
        {CURE-worse-15}
\bibfield{author}{\bibinfo{person}{Edward~K. Smith}, \bibinfo{person}{Earl~T.
  Barr}, \bibinfo{person}{Claire Le~Goues}, {and} \bibinfo{person}{Yuriy
  Brun}.} \bibinfo{year}{2015}\natexlab{}.
\newblock \showarticletitle{Is the Cure Worse Than the Disease? Overfitting in
  Automated Program Repair}. In \bibinfo{booktitle}{\emph{Proceedings of the
  2015 10th Joint Meeting on Foundations of Software Engineering}}
  \emph{(\bibinfo{series}{ESEC/FSE 2015})}.
\newblock


\bibitem[\protect\citeauthoryear{Tian, Liu, Kabor{\'e}, Koyuncu, Li, Klein, and
  Bissyand{\'e}}{Tian et~al\mbox{.}}{2020}]%
        {TianASE20}
\bibfield{author}{\bibinfo{person}{Haoye Tian}, \bibinfo{person}{Kui Liu},
  \bibinfo{person}{Abdoul~Kader Kabor{\'e}}, \bibinfo{person}{Anil Koyuncu},
  \bibinfo{person}{Li Li}, \bibinfo{person}{Jacques Klein}, {and}
  \bibinfo{person}{Tegawend{\'e}~F. Bissyand{\'e}}.}
  \bibinfo{year}{2020}\natexlab{}.
\newblock \showarticletitle{Evaluating Representation Learning of Code Changes
  for Predicting Patch Correctness in Program Repair}. In
  \bibinfo{booktitle}{\emph{Proceedings of the 35th IEEE/ACM International
  Conference on Automated Software Engineering}}. \bibinfo{publisher}{{IEEE}},
  \bibinfo{pages}{981--992}.
\newblock
\urldef\tempurl%
\url{https://doi.org/10.1145/3324884.3416532}
\showDOI{\tempurl}


\bibitem[\protect\citeauthoryear{Tufano, Pantiuchina, Watson, Bavota, and
  Poshyvanyk}{Tufano et~al\mbox{.}}{2019a}]%
        {Tufano-ICSE19}
\bibfield{author}{\bibinfo{person}{Michele Tufano}, \bibinfo{person}{Jevgenija
  Pantiuchina}, \bibinfo{person}{Cody Watson}, \bibinfo{person}{Gabriele
  Bavota}, {and} \bibinfo{person}{Denys Poshyvanyk}.}
  \bibinfo{year}{2019}\natexlab{a}.
\newblock \showarticletitle{On Learning Meaningful Code Changes via Neural
  Machine Translation}. In \bibinfo{booktitle}{\emph{Proceedings of the 41st
  International Conference on Software Engineering}} (Montreal, Quebec, Canada)
  \emph{(\bibinfo{series}{ICSE '19})}. \bibinfo{publisher}{IEEE Press},
  \bibinfo{pages}{25–36}.
\newblock
\urldef\tempurl%
\url{https://doi.org/10.1109/ICSE.2019.00021}
\showDOI{\tempurl}


\bibitem[\protect\citeauthoryear{Tufano, Watson, Bavota, Penta, White, and
  Poshyvanyk}{Tufano et~al\mbox{.}}{2019b}]%
        {Tufano-tse19}
\bibfield{author}{\bibinfo{person}{Michele Tufano}, \bibinfo{person}{Cody
  Watson}, \bibinfo{person}{Gabriele Bavota}, \bibinfo{person}{Massimiliano~Di
  Penta}, \bibinfo{person}{Martin White}, {and} \bibinfo{person}{Denys
  Poshyvanyk}.} \bibinfo{year}{2019}\natexlab{b}.
\newblock \showarticletitle{An Empirical Study on Learning Bug-Fixing Patches
  in the Wild via Neural Machine Translation}.
\newblock \bibinfo{journal}{\emph{ACM Trans. Softw. Eng. Methodol.}}
  \bibinfo{volume}{28}, \bibinfo{number}{4}, Article \bibinfo{articleno}{19}
  (\bibinfo{date}{Sept.} \bibinfo{year}{2019}), \bibinfo{numpages}{29}~pages.
\newblock
\showISSN{1049-331X}
\urldef\tempurl%
\url{https://doi.org/10.1145/3340544}
\showDOI{\tempurl}


\bibitem[\protect\citeauthoryear{Vaswani, Shazeer, Parmar, Uszkoreit, Jones,
  Gomez, Kaiser, and Polosukhin}{Vaswani et~al\mbox{.}}{2017a}]%
        {attention-all-you-need}
\bibfield{author}{\bibinfo{person}{Ashish Vaswani}, \bibinfo{person}{Noam
  Shazeer}, \bibinfo{person}{Niki Parmar}, \bibinfo{person}{Jakob Uszkoreit},
  \bibinfo{person}{Llion Jones}, \bibinfo{person}{Aidan~N Gomez},
  \bibinfo{person}{\L~ukasz Kaiser}, {and} \bibinfo{person}{Illia Polosukhin}.}
  \bibinfo{year}{2017}\natexlab{a}.
\newblock \showarticletitle{Attention is All you Need}. In
  \bibinfo{booktitle}{\emph{Advances in Neural Information Processing
  Systems}}, Vol.~\bibinfo{volume}{30}.
\newblock


\bibitem[\protect\citeauthoryear{Vaswani, Shazeer, Parmar, Uszkoreit, Jones,
  Gomez, Kaiser, and Polosukhin}{Vaswani et~al\mbox{.}}{2017b}]%
        {Vaswani2021Transformer}
\bibfield{author}{\bibinfo{person}{Ashish Vaswani}, \bibinfo{person}{Noam
  Shazeer}, \bibinfo{person}{Niki Parmar}, \bibinfo{person}{Jakob Uszkoreit},
  \bibinfo{person}{Llion Jones}, \bibinfo{person}{Aidan~N. Gomez},
  \bibinfo{person}{undefinedukasz Kaiser}, {and} \bibinfo{person}{Illia
  Polosukhin}.} \bibinfo{year}{2017}\natexlab{b}.
\newblock \showarticletitle{Attention is All You Need}. In
  \bibinfo{booktitle}{\emph{Proceedings of the 31st International Conference on
  Neural Information Processing Systems}} (Long Beach, California, USA)
  \emph{(\bibinfo{series}{NIPS'17})}. \bibinfo{publisher}{Curran Associates
  Inc.}, \bibinfo{address}{Red Hook, NY, USA}, \bibinfo{pages}{6000–6010}.
\newblock
\showISBNx{9781510860964}


\bibitem[\protect\citeauthoryear{Wang, Singh, and Su}{Wang
  et~al\mbox{.}}{2018a}]%
        {wang2018dynamic}
\bibfield{author}{\bibinfo{person}{Ke Wang}, \bibinfo{person}{Rishabh Singh},
  {and} \bibinfo{person}{Zhendong Su}.} \bibinfo{year}{2018}\natexlab{a}.
\newblock \showarticletitle{Dynamic Neural Program Embedding for Program
  Repair} \emph{(\bibinfo{series}{ICLR})}.
\newblock


\bibitem[\protect\citeauthoryear{Wang, Singh, and Su}{Wang
  et~al\mbox{.}}{2018b}]%
        {wang-sar}
\bibfield{author}{\bibinfo{person}{Ke Wang}, \bibinfo{person}{Rishabh Singh},
  {and} \bibinfo{person}{Zhendong Su}.} \bibinfo{year}{2018}\natexlab{b}.
\newblock \showarticletitle{Search, Align, and Repair: Data-Driven Feedback
  Generation for Introductory Programming Exercises}.
\newblock \bibinfo{journal}{\emph{SIGPLAN Not.}} \bibinfo{volume}{53},
  \bibinfo{number}{4} (\bibinfo{date}{June} \bibinfo{year}{2018}),
  \bibinfo{pages}{481–495}.
\newblock
\showISSN{0362-1340}
\urldef\tempurl%
\url{https://doi.org/10.1145/3296979.3192384}
\showDOI{\tempurl}


\bibitem[\protect\citeauthoryear{Wang and Su}{Wang and Su}{2020}]%
        {embed-ke-PLDI20}
\bibfield{author}{\bibinfo{person}{Ke Wang} {and} \bibinfo{person}{Zhendong
  Su}.} \bibinfo{year}{2020}\natexlab{}.
\newblock \showarticletitle{Blended, Precise Semantic Program Embeddings}. In
  \bibinfo{booktitle}{\emph{Proceedings of the 41st ACM SIGPLAN Conference on
  Programming Language Design and Implementation}} (London, UK)
  \emph{(\bibinfo{series}{PLDI 2020})}. \bibinfo{publisher}{Association for
  Computing Machinery}, \bibinfo{address}{New York, NY, USA},
  \bibinfo{pages}{121–134}.
\newblock
\showISBNx{9781450376136}
\urldef\tempurl%
\url{https://doi.org/10.1145/3385412.3385999}
\showDOI{\tempurl}


\bibitem[\protect\citeauthoryear{Wang, Wen, Lin, Wu, Qin, Zou, Mao, and
  Jin}{Wang et~al\mbox{.}}{2020}]%
        {ASE20Wang}
\bibfield{author}{\bibinfo{person}{Shangwen Wang}, \bibinfo{person}{Ming Wen},
  \bibinfo{person}{Bo Lin}, \bibinfo{person}{Hongjun Wu},
  \bibinfo{person}{Yihao Qin}, \bibinfo{person}{Deqing Zou},
  \bibinfo{person}{Xiaoguang Mao}, {and} \bibinfo{person}{Hai Jin}.}
  \bibinfo{year}{2020}\natexlab{}.
\newblock \showarticletitle{Automated Patch Correctness Assessment: How Far are
  We?}. In \bibinfo{booktitle}{\emph{Proceedings of the 35th International
  Conference on Automated Software Engineering (ASE)}}. ACM.
\newblock


\bibitem[\protect\citeauthoryear{Wen, Chen, Wu, Hao, and Cheung}{Wen
  et~al\mbox{.}}{2018}]%
        {capgen-ICSE18}
\bibfield{author}{\bibinfo{person}{Ming Wen}, \bibinfo{person}{Junjie Chen},
  \bibinfo{person}{Rongxin Wu}, \bibinfo{person}{Dan Hao}, {and}
  \bibinfo{person}{Shing-Chi Cheung}.} \bibinfo{year}{2018}\natexlab{}.
\newblock \showarticletitle{Context-Aware Patch Generation for Better Automated
  Program Repair}. In \bibinfo{booktitle}{\emph{Proceedings of the 40th
  International Conference on Software Engineering}}
  \emph{(\bibinfo{series}{ICSE '18})}.
\newblock


\bibitem[\protect\citeauthoryear{Xin and Reiss}{Xin and Reiss}{2019}]%
        {sharpFix}
\bibfield{author}{\bibinfo{person}{Qi Xin} {and} \bibinfo{person}{Steven
  Reiss}.} \bibinfo{year}{2019}\natexlab{}.
\newblock \showarticletitle{Better Code Search and Reuse for Better Program
  Repair}. In \bibinfo{booktitle}{\emph{2019 IEEE/ACM International Workshop on
  Genetic Improvement (GI)}}. \bibinfo{pages}{10--17}.
\newblock
\urldef\tempurl%
\url{https://doi.org/10.1109/GI.2019.00012}
\showDOI{\tempurl}


\bibitem[\protect\citeauthoryear{Xiong, Wang, Yan, Zhang, Han, Huang, and
  Zhang}{Xiong et~al\mbox{.}}{2017}]%
        {acs}
\bibfield{author}{\bibinfo{person}{Yingfei Xiong}, \bibinfo{person}{Jie Wang},
  \bibinfo{person}{Runfa Yan}, \bibinfo{person}{Jiachen Zhang},
  \bibinfo{person}{Shi Han}, \bibinfo{person}{Gang Huang}, {and}
  \bibinfo{person}{Lu Zhang}.} \bibinfo{year}{2017}\natexlab{}.
\newblock \showarticletitle{Precise Condition Synthesis for Program Repair}. In
  \bibinfo{booktitle}{\emph{Proceedings of the 39th International Conference on
  Software Engineering}} (Buenos Aires, Argentina) \emph{(\bibinfo{series}{ICSE
  '17})}. \bibinfo{publisher}{IEEE Press}, \bibinfo{pages}{416–426}.
\newblock
\showISBNx{9781538638682}
\urldef\tempurl%
\url{https://doi.org/10.1109/ICSE.2017.45}
\showDOI{\tempurl}


\bibitem[\protect\citeauthoryear{Xuan, Martinez, Demarco, Clément, Lamelas,
  Durieux, Le~Berre, and Monperrus}{Xuan et~al\mbox{.}}{2016}]%
        {nopol}
\bibfield{author}{\bibinfo{person}{Jifeng Xuan}, \bibinfo{person}{Matias
  Martinez}, \bibinfo{person}{Favio Demarco}, \bibinfo{person}{Maxime
  Clément}, \bibinfo{person}{Sebastian Lamelas}, \bibinfo{person}{Thomas
  Durieux}, \bibinfo{person}{Daniel Le~Berre}, {and} \bibinfo{person}{Martin
  Monperrus}.} \bibinfo{year}{2016}\natexlab{}.
\newblock \showarticletitle{Nopol: Automatic Repair of Conditional Statement
  Bugs in Java Programs}.
\newblock \bibinfo{journal}{\emph{IEEE Transactions on Software Engineering}}
  (\bibinfo{year}{2016}).
\newblock


\bibitem[\protect\citeauthoryear{Ye, Gu, Martinez, Durieux, and Monperrus}{Ye
  et~al\mbox{.}}{2021a}]%
        {ODS}
\bibfield{author}{\bibinfo{person}{He Ye}, \bibinfo{person}{Jian Gu},
  \bibinfo{person}{Matias Martinez}, \bibinfo{person}{Thomas Durieux}, {and}
  \bibinfo{person}{Martin Monperrus}.} \bibinfo{year}{2021}\natexlab{a}.
\newblock \showarticletitle{Automated Classification of Overfitting Patches
  with Statically Extracted Code Features}.
\newblock \bibinfo{journal}{\emph{IEEE Transactions on Software Engineering}}
  (\bibinfo{year}{2021}).
\newblock
\urldef\tempurl%
\url{https://doi.org/10.1109/tse.2021.3071750}
\showDOI{\tempurl}


\bibitem[\protect\citeauthoryear{Ye, Martinez, Durieux, and Monperrus}{Ye
  et~al\mbox{.}}{2021c}]%
        {quixbugs-jss}
\bibfield{author}{\bibinfo{person}{He Ye}, \bibinfo{person}{Matias Martinez},
  \bibinfo{person}{Thomas Durieux}, {and} \bibinfo{person}{Martin Monperrus}.}
  \bibinfo{year}{2021}\natexlab{c}.
\newblock \showarticletitle{A comprehensive study of automatic program repair
  on the QuixBugs benchmark}.
\newblock \bibinfo{journal}{\emph{Journal of Systems and Software}}
  \bibinfo{volume}{171} (\bibinfo{year}{2021}), \bibinfo{pages}{110825}.
\newblock
\showISSN{0164-1212}
\urldef\tempurl%
\url{https://doi.org/10.1016/j.jss.2020.110825}
\showDOI{\tempurl}


\bibitem[\protect\citeauthoryear{Ye, Martinez, Luo, Zhang, and Monperrus}{Ye
  et~al\mbox{.}}{2022}]%
        {selfapr}
\bibfield{author}{\bibinfo{person}{He Ye}, \bibinfo{person}{Matias Martinez},
  \bibinfo{person}{Xiapu Luo}, \bibinfo{person}{Tao Zhang}, {and}
  \bibinfo{person}{Martin Monperrus}.} \bibinfo{year}{2022}\natexlab{}.
\newblock \showarticletitle{SelfAPR: Self-supervised Program Repair with Test
  Execution Diagnostics}. In \bibinfo{booktitle}{\emph{arXiv}}.
\newblock


\bibitem[\protect\citeauthoryear{Ye, Martinez, and Monperrus}{Ye
  et~al\mbox{.}}{2021b}]%
        {drr}
\bibfield{author}{\bibinfo{person}{He Ye}, \bibinfo{person}{Matias Martinez},
  {and} \bibinfo{person}{Martin Monperrus}.} \bibinfo{year}{2021}\natexlab{b}.
\newblock \showarticletitle{Automated patch assessment for program repair at
  scale}.
\newblock \bibinfo{journal}{\emph{Empirical Software Engineering}}
  \bibinfo{volume}{26}, \bibinfo{number}{2} (\bibinfo{year}{2021}),
  \bibinfo{pages}{20}.
\newblock
\showISBNx{1573-7616}
\urldef\tempurl%
\url{https://doi.org/10.1007/s10664-020-09920-w}
\showDOI{\tempurl}


\bibitem[\protect\citeauthoryear{Yu, Zhang, Wang, and Yu}{Yu
  et~al\mbox{.}}{2017}]%
        {seqGan-aaai17}
\bibfield{author}{\bibinfo{person}{Lantao Yu}, \bibinfo{person}{Weinan Zhang},
  \bibinfo{person}{Jun Wang}, {and} \bibinfo{person}{Yong Yu}.}
  \bibinfo{year}{2017}\natexlab{}.
\newblock \showarticletitle{SeqGAN: Sequence Generative Adversarial Nets with
  Policy Gradient}. In \bibinfo{booktitle}{\emph{Proceedings of the
  Thirty-First AAAI Conference on Artificial Intelligence}} (San Francisco,
  California, USA) \emph{(\bibinfo{series}{AAAI'17})}. \bibinfo{publisher}{AAAI
  Press}, \bibinfo{pages}{2852–2858}.
\newblock


\bibitem[\protect\citeauthoryear{Yu, Martinez, Danglot, Durieux, and
  Monperrus}{Yu et~al\mbox{.}}{2018}]%
        {zhongxing-EMSE18}
\bibfield{author}{\bibinfo{person}{Zhongxing Yu}, \bibinfo{person}{Matias
  Martinez}, \bibinfo{person}{Benjamin Danglot}, \bibinfo{person}{Thomas
  Durieux}, {and} \bibinfo{person}{Martin Monperrus}.}
  \bibinfo{year}{2018}\natexlab{}.
\newblock \showarticletitle{Alleviating patch overfitting with automatic test
  generation: a study of feasibility and effectiveness for the Nopol repair
  system}.
\newblock \bibinfo{journal}{\emph{Empirical Software Engineering}}
  (\bibinfo{year}{2018}).
\newblock


\bibitem[\protect\citeauthoryear{Yuan and Banzhaf}{Yuan and Banzhaf}{2018}]%
        {Yuan2017ARJAAR}
\bibfield{author}{\bibinfo{person}{Yuan Yuan} {and} \bibinfo{person}{Wolfgang
  Banzhaf}.} \bibinfo{year}{2018}\natexlab{}.
\newblock \showarticletitle{ARJA: Automated Repair of Java Programs via
  Multi-Objective Genetic Programming}. In \bibinfo{booktitle}{\emph{IEEE
  Transactions on Software Engineering}}.
\newblock


\bibitem[\protect\citeauthoryear{Zhang, Feng, Meng, You, and Liu}{Zhang
  et~al\mbox{.}}{2019}]%
        {NMT-bridging}
\bibfield{author}{\bibinfo{person}{Wen Zhang}, \bibinfo{person}{Yang Feng},
  \bibinfo{person}{Fandong Meng}, \bibinfo{person}{Di You}, {and}
  \bibinfo{person}{Qun Liu}.} \bibinfo{year}{2019}\natexlab{}.
\newblock \showarticletitle{Bridging the Gap between Training and Inference for
  Neural Machine Translation}. In \bibinfo{booktitle}{\emph{Proceedings of the
  57th Annual Meeting of the Association for Computational Linguistics}}.
  \bibinfo{publisher}{Association for Computational Linguistics},
  \bibinfo{address}{Florence, Italy}, \bibinfo{pages}{4334--4343}.
\newblock
\urldef\tempurl%
\url{https://doi.org/10.18653/v1/P19-1426}
\showDOI{\tempurl}


\bibitem[\protect\citeauthoryear{Zhao, Vogel, and Ahmed}{Zhao
  et~al\mbox{.}}{2019}]%
        {lossscale}
\bibfield{author}{\bibinfo{person}{Ruizhe Zhao}, \bibinfo{person}{Brian Vogel},
  {and} \bibinfo{person}{Tanvir Ahmed}.} \bibinfo{year}{2019}\natexlab{}.
\newblock \bibinfo{title}{Adaptive Loss Scaling for Mixed Precision Training}.
\newblock
\newblock
\showeprint[arxiv]{1910.12385}


\bibitem[\protect\citeauthoryear{Zhu, Sun, Xiao, Zhang, Yuan, Xiong, and
  Zhang}{Zhu et~al\mbox{.}}{2021}]%
        {Recoder}
\bibfield{author}{\bibinfo{person}{Qihao Zhu}, \bibinfo{person}{Zeyu Sun},
  \bibinfo{person}{Yuan-an Xiao}, \bibinfo{person}{Wenjie Zhang},
  \bibinfo{person}{Kang Yuan}, \bibinfo{person}{Yingfei Xiong}, {and}
  \bibinfo{person}{Lu Zhang}.} \bibinfo{year}{2021}\natexlab{}.
\newblock \showarticletitle{A Syntax-Guided Edit Decoder for Neural Program
  Repair}. In \bibinfo{booktitle}{\emph{Proceedings of the 29th ACM Joint
  Meeting on European Software Engineering Conference and Symposium on the
  Foundations of Software Engineering}} (Athens, Greece)
  \emph{(\bibinfo{series}{ESEC/FSE 2021})}. \bibinfo{publisher}{Association for
  Computing Machinery}, \bibinfo{address}{New York, NY, USA},
  \bibinfo{pages}{341–353}.
\newblock
\showISBNx{9781450385626}
\urldef\tempurl%
\url{https://doi.org/10.1145/3468264.3468544}
\showDOI{\tempurl}


\end{thebibliography}

\end{document}